# Learning with Digital Agents: An Analysis based on the Activity Theory

**Authors:**

Mateusz Dolata, University of Zurich, Binzmuehlestrasse 14, CH-8050 Zurich, Switzerland, dolata@ifi.uzh.ch (corresponding author)

Dzmitry Katsiuba, University of Zurich, Binzmuehlestrasse 14, CH-8050 Zurich, Switzerland, katsiuba@ifi.uzh.ch

Natalie Wellnhammer, BSI Business Systems Integration AG, Taefernweg 1, CH-5405 Baden, Switzerland, natalie.wellnhammer@bsi-software.com

Gerhard Schwabe, University of Zurich, Binzmuehlestrasse 14, CH-8050 Zurich, Switzerland, schwabe@ifi.uzh.ch

**Abstract:**

Digital agents are considered a general-purpose technology. They spread quickly in private and organizational contexts, including education. Yet, research lacks a conceptual framing to describe interaction with such agents in a holistic manner. While focusing on the interaction with a pedagogical agent, i.e., a digital agent capable of natural-language interaction with a learner, we propose a model of learning activity based on activity theory. We use this model and a review of prior research on digital agents in education to analyze how various characteristics of the activity, including features of a pedagogical agent or learner, influence learning outcomes. The analysis leads to identification of IS research directions and guidance for developers of pedagogical agents and digital agents in general. We conclude by extending the activity theory-based model beyond the context of education and show how it helps designers and researchers ask the right questions when creating a digital agent.

## Bio Sketches

**Mateusz Dolata**, born 1988, currently hired as a full-time postdoctoral researcher at the University of Zurich, Department of Informatics. His interest spans co-located collaboration in IT-supported settings, responsible application of artificial intelligence for the common good, and application of digital agents in organizational settings. His articles open the black box of work practices in human-computer assemblages while questioning scientific and public discourses that accompany the proliferation of modern technologies.

**Dzmitry Katsiuba**, born 1987, currently pursuing his PhD as a full-time research assistant at the University of Zurich, Department of Informatics. He has a background in information systems, education, and geography. His primary research interest is the collaboration between artificial intelligence and human workers in organizational context. He co-authored articles on the use of natural language processing and artificial intelligence in professional work.

**Natalie Wellnhammer**, born 1994, alumna of the University of Zurich, Department of Informatics, currently pursuing her professional career as a software engineer at BSI Business Systems Integration AG. She initiated the collaborative effort to study pedagogical agents and was the first author of the HICSS article which preceded the current study. She is interested in ways to bridge pedagogy and professional work.

**Gerhard Schwabe**, born 1962, full professor at the University of Zurich, Department of Informatics. He specializes in information management and collaborative technologies. In books, journal articles, and conference publications he focuses on collaboration in workshops, city councils, learning, advice giving, and in digital platforms. His research focuses on the impact, application potential and management of novel technologies such as tangible interfaces, blockchain, artificial intelligence, and human-robot collaboration.

## INTRODUCTION

Digital agents (DA) are becoming ubiquitous in private and organizational contexts. They are digital systems offering a virtual character that interacts with the user by any combination of visual, symbolic, and natural-language communication [58,167,197]. The virtual character might be explicit, possessing a name and a visual representation, or implicit, emerging through its actions [155,197]. The interaction with a DA imitates interaction with another human, though perhaps limited by available media (e.g., text-based chat vs. speech vs. visual input and output) and abilities of the DA (e.g., guidance vs. accepting and answering predefined input vs. accepting and answering free input). Other terms for DA include virtual agent/assistant or conversational agent. IS research proposed numerous applications in which DA are used to reduce cognitive load, enhance customer experience, or improve health outcomes and knowledge acquisition compared to situations without a DA [14,83,127,133,136,167,189]. Given the development of natural language understanding, speech processing, and multimodal interfaces, DA will pervade all major areas of human activity soon.

Reports on the advantages of DAs cumulate. However, it remains an open question how DA accomplish those improvements, i.e., what characteristics of the DA, the situation, or the user, as well as what social and cognitive processes on the user's side are involved in yielding enhanced outcomes. DA trigger social and emotional processing apart from a goal-oriented response in the user [139,197], so conventional models of human-technology interaction and existing design guidance are of limited use. At the same time, the design space for DA is larger than in conventional user interfaces: each word, utterance, or atomic behavior of a DA can generate an undesired reaction from the user [83], and might need careful consideration from the designer. DA developers are frequently overwhelmed by this task and lack specific guidance on how to design an artefact that potentially will be considered a social entity by the user [167]. IS needs both theories of the interaction between users and DA for a better understanding of what happens in such settings and, ultimately, to render the much-needed guidance. In this manuscript, we attend to an application domain with a long tradition of employing DA: education. We use it as a context for theorizing the interaction between a human user and a DA, while identifying challenges relevant to IS more generally.

Use of DA in education is a controversial topic. Past research shows that DAs have positive impacts on the cognitive and emotional aspects of learning [9], but also generate concerns regarding psychological welfare, efficiency, or usability [172]. In fact, some studies directly contradict the positive ones or are inconclusive as to whether DA improve or impede learning [9,113,172]. This highlights the challenges of development and rollout of DAs. IS recently joined the discourse on pedagogical use of DAs by identifying potential for DA in IS education, studying the organizational consequences of using DA in higher education and vocational training, and empowering teachers to integrate DA in their teaching [36,88,106,108,158,185,190,191,193]. This research indicates that application of DAs in education not only impacts the learning outcomes and experiences of the student, but bears potential for facilitating life-long learning and upskilling of employees, as well as for improving the practice of teaching and learning [191,192,195]. We see e-learning and use of technology in education as important research and application area for IS. This article positions the research on use of DA in education at the intersection of the individual, the organization, and the technology, making it even more relevant for IS.

Throughout this article, we refer to DAs used in education as *pedagogical agents* (PA). A PA is a digital agent capable of communication in natural language (written, spoken, or sign language) designed to help a human learner improve their knowledge or skills. A DA might be represented as an avatar within an intelligent tutoring system (ITS) [4,34,113] or an e-learning environment [106], be a standalone application using popular technologies like Alexa [190,193], or even be a social robot [9]. Numerous studies offer insights on how PAs impact learning outcomes and related constructs like motivation or self-efficacy. PAs employed in those studies differ in terms of their capabilities, features, interaction principles, roles they play, and the context they are designed for. Also, the studies differ in terms of learner category (primary school children vs. university students), topic or subject (physics vs. informatics), or material used (videos and visualizations vs. speech only). Finally, the studies are spread across disciplinary boundaries and use various terms to describe PA. Overall, it might be hard to find and differentiate between studies with higher and lower relevance when designing a PA for an educational setting. However, a comprehensive and yet detailed summary of the literature is necessary since a single aspect of a PA can have significant impact on the learner and the learning outcome [126,193].

Even though meta-studies and systematic reviews are available, they frequently are of little use to inform the process of designing, developing, and deploying a PA. On the one hand, many meta-studies try to answer the question whether PAs help improve learning or not [6,118,166,196]. This approach ignores the relevant differences among PAs. On the other hand, some meta-studies focus on a singular design aspect, e.g., gesturing, and summarize how variations in this single aspect might impact success in learning [19,42,85]. Those studies might inform singular design decisions but not explain how those granular decisions fit together. We claim that PAs are elements of a complex, socio-technical environment, with various aspects interacting with each other, thus requiring a comprehensive approach. Furthermore, many design aspects, like a PA's adaptivity to the learner, were not attended by meta-studies yet. Even though some comprehensive meta-studies are available, they focus on bibliographic information such as the outlet's domain [26,200] or characteristics of the studies (e.g., test item format, pre-treatment differences, study duration, sample size) [113], rather than on describing the PA or its usage. Information provided in such summaries is of interest to academic audiences and might yield research agendas but fall short on design guidance. Finally, many reviews cover a subset of available studies by focusing on an application domain, e.g., primary and secondary school, [117,196], or type of PA, e.g., chatbot, ITS, or social robots [9,96,194]. For one thing, they attend to the mainstream research while leaving out niche developments like PA for vocational training. For another, they reinforce disciplinary boundaries by using framing or terminology of a specific domain, e.g., "ITS" for education technology vs. "chatbot" for computer science. Even though the terms might describe very similar or identical systems, some meta-studies stay within disciplinary borders. Those disciplinary distinctions are irrelevant for PA designers. Instead, they require an overview of the challenges they will encounter when designing a PA and an understanding of the key factors to consider as they attempt to improve learning outcomes by means of a PA. Overall, the existing reviews and meta-studies neither provide a comprehensive theoretical framework for designers when creating PA or to inform comprehensive analysis of learning with PA, nor do they yield a consistent picture of which design aspects might require careful context-dependent design or further research, or are controversial.

The current article aims at summarizing the results of PA scholarship through a systematic literature review. It provides an analysis of features or interaction capabilities of a PA that moderate the agent's impact on the learning outcome. Following this objective, we ask two research questions:

RQ1: *How can we conceptualize a pedagogical agent as an entity within an activity system?*

RQ2: *How does the design of a pedagogical agent impact learning outcomes?*

The answer to those questions should guide PA designers. Teachers and schools benefit from an overview of important relevant aspects when deciding on the use of PAs in specific contexts. Educational scientists get an updated systematic overview of PA research. Finally, IS researchers get insight into a research area that attracted more attention recently [88,191] but has a 25-year tradition in education. Given that DA became a core IS topic only in the recent years, many IS researchers might be surprised that an adjacent discipline first rendered real-world applications relying on this concept in the 1990s. We strongly advocate learning from the experience and insights produced by this research to inform the design of DA for other domains including commerce, services, health, or public administration.

This manuscript introduces an activity-theoretical framework of *learning with a pedagogical agent* (LPA) to drive and structure the analysis. In fact, the underlying literature research shows that theorizing about PA and its effect is still rudimentary. The studies resort to theories from either pedagogy or subdisciplines of computer science, but a theoretical lens uniting those disciplines is absent. The proposed framework, LPA, builds bridges across disciplinary boundaries. It relies on activity theory (AT), a theoretical underpinning with extended tradition in pedagogy, practice-oriented computing, and IS. It sees the interaction between an individual learner[1] and a PA as embedded in a larger, contextualized activity. By positioning the PA as a perceived subject, it proposes an update to AT and makes it applicable to interaction between human and non-human agents. This bears potential not only for understanding PA-supported learning, but also for a new theoretical perspective on activities involving DA independent of their context. We develop a model to describe activities involving a DA, the *'activity with a digital agent' model* (ADAM), to abstract it from the application in education.

The literature analysis driven by this framework identifies research potentials, highlighting the most challenging design decisions. The review demonstrates that the choice of media and visual or bodily presentation of the agent were extensively addressed in the past, while the social, material, and technical contexts of LPA remain underexplored. On the one hand, this raises research questions for IS given its interest in understanding the interdependencies between the technology, the context of use, and the social and organizational impact. On the other hand, it pinpoints areas that require specific attention from PA designers: aspects that have not been researched yet in detail, yielded contradictory evidence, or are subject to moderation or interdependencies. By accentuating the holistic nature of LPA, the article provides a tool for understanding the synergies and contradictions from a new perspective.

---

[1] While we acknowledge that many learners could interact with a DA or a group of DAs, e.g., when DA acts as a classroom teacher or a group member, this manuscript focuses on the single-learner scenario. This is the simplest and the best-researched application for PAs: only few studies attend to interaction between a PA and multiple learners. The chosen focus spans various scenarios, e.g., exam preparation, homework, practicing work skills, or individual study during a class. Still, we see a school, a university, or a course as the organizational context, and the community of teachers, assistants, and learners as the social context of the learning activity.

The manuscript unfolds as follows. First, it uses AT to derive the theoretical perspective on LPA. Second, it uses this framework to systematically analyze studies that link features of PA or general characteristics of the situation to the learning outcome. Thereby it focuses on the manipulated and controlled aspects (e.g., design decisions) and how those influence learning outcomes. The manuscript provides a detailed overview of the methodology used for selection and analysis of the literature, as well as the results. Third, the study reflects on the implication of the results for the design of PA and, in more general terms, for the research about DA. Finally, it comments on the applicability of the insights.

## PEDAGOGICAL AGENTS WITHIN AN ACTIVITY SYSTEM

To establish a framework for the analysis of PA, we turn to AT. AT has been applied extensively in pedagogy since its inception and recently found supporters in the IS. We introduce the core concepts of AT, including their source and primary meaning, and proceed to open questions emerging in pedagogy and IS. We conclude by developing an activity system for *learning with a pedagogical agent*.

### Activity Theory in Education

The origins of AT go back to Russian psychology of the 1920s –1930s. In the early works of Vygotsky, the founder of AT, the theory was used to explain development of the human mind, especially the acquisition of language or scientific concepts by children [41,102,115,183,184]. The basic mechanism identified by Vygotsky was the *mediation,* which enables the human *subject* to approach an external *object* by means of psychological and material *tools* [41,183]. Accordingly, an *activity* is an interaction between the subject and the object mediated by a set of tools and oriented at transforming the object [41,114] (cf. Figure 5a). Activities are distinguished by their objects; the intention to transform the object motivates activity [114]. While material tools form extensions of a human body and provide new embodied stimuli (e.g., a hammer in use), psychological tools, including language or sign systems, mediate thoughts [115]. Thereby material tools can exercise a reverse, reciprocal action, i.e., they are not simply employed by the subject to the objects in the world, but they also enable the subject to develop new psychological functions [41,102,115,183] – to learn. The AT was formulated as a way to overcome the dichotomy between material and psychological experiences, as well as the contradiction between internal and external stimuli in an individual [102].

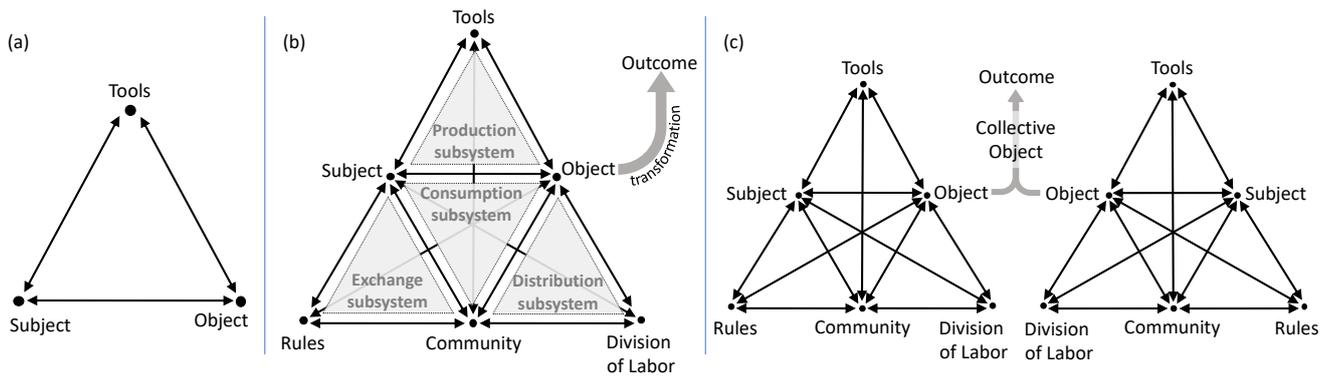

*Figure 1. Three generations of activity theory: (a) an activity system according to the first generation AT as devised by Vygotsky [183], (b) second generation AT as proposed by Leontyev [120] and represented by Engeström [68] including the subsystems of an activity system, (c) third generation AT as proposed by Engeström [68,70,72] for describing interdependent activity systems.*

Leontyev, a student and colleague of Vygotsky, extended the AT to the collective aspects of activity. Accordingly, development of the human mind is not limited to an individual's interaction with an external object, but happens in a social context that gives meaning to the activities of members of the community [102,103,120]. Also, many tools emerge in a social interaction (e.g., language, signs) and only later get appropriated for individual use [102,120]. Human activities can be very complex, requiring many separate steps and various tools, such that proficiency in each aspect cannot be maintained by a single individual [102]. This leads to the emergence of *division of labor* and *rules and norms* that help maintain the distribution of activities between the subject and the *community* [120]. Leontyev's take on AT helped in overcoming the dichotomy between individual and collective activities, including situated collective activity as a core aspect of human development and learning, showing that improvement within an individual is not reducible to detached individual actions but requires social action [98].

Even though Leontyev explained the role of community in human activity, only decades later, did Engeström formulate an activity system as a framework for analysis of six interconnected components: *subject, object, tool, community, rules,* and *division of labor* [68,69,71]. Additionally, Engeström [68] identifies four higher-order functions within the primary activity system: *production, distribution, exchange,* and *consumption*, that arise from mutual relations between the components of each subsystem (cf. Figure 5b) [68,89,98]. Production is oriented at the outcome of the activity system and leads to the development of an object that meets the subject's need by means of available tools. Distribution ties the object of the activity to the community by dividing it among its members. Exchange regulates the activities of the system in terms of personal needs of the subject and the members of the community. Finally, consumption describes the collaborative actions of the subject and the community on the object, while resources are consumed to transform the object. Overall, an activity system describes components and functions involved in transforming an object to meet the needs of a subject.

The works of Engeström mark the beginning of popularization of the AT in Western research on pedagogy and education [86,154]. Engeström's work on adult learning and organizational change has significant impacts on the conceptual structure of AT. He moved the focus of AT towards interconnected ensembles of activity systems which enabled the study of interrelations among multiple activities involving multiple subjects [71,103] (cf. Figure 5c), e.g., as in a situation when a patient is treated in a hospital first (one activity system) but then the local doctor (another activity system) takes care of her recovery. Based on empirical studies, he then identified contradictions and conflicts between and within activity systems as the reason behind their evolution, e.g., as in a situation when a new tool introduced into an activity system contradicts existing social rules and norms [69,102]. This mechanism provides conceptualization for collective learning. Engeström's concepts were widely adopted in pedagogy and education as a tool for analysis and intervention in schools and other educational settings [82,142,154].

A full review of AT in pedagogy goes beyond the scope of this article, but the above summary brings some limitations to the surface. First, AT relies on differentiation between social agents (e.g., subject, community) and non-social entities (e.g., tools). Given the increasing hybridization of technologies and humans in the context of education (social networks, massive online open courses, PA), it remains uncertain whether and how AT can accommodate those developments [103,160] and what component of the AT can be associated with non-living entities that behave socially and can be perceived by humans as behaving in accordance with an internal, or even externally-imposed, intention [143,160,164]. Second, AT (like other general-purpose theories of human activity) relies on a black-box concept of the *tool* [103] – its singular elements or features and their impact on producing the desired outcome remain unpacked. The pedagogical AT literature contains various examples of tools including computer-based learning software [81], conversations, questions, or group session [175], and textbooks, computers, or even human agents like professors [98]. This short list readily illustrates how the understanding of tools differs across studies and sub-disciplines. Unpacking various technologies and analyzing their properties is central to understanding how they impact the learning outcome [103]. Despite these limitations, AT has proven to be a very useful framework for study and development of education: it acknowledges that learning occurs most naturally and meaningfully in the context of an activity [98], in which the cognition is distributed across people and artefacts [178], and it paved the way to successful and effective interventions in educational and organizational contexts [67].

Third, AT originally framed the acquisition of new psychological functions as a consequence of reciprocal action between the subject and the object, mediated via the tool [102,114,184], but the instantiations of those concepts vary significantly. For instance, there is an ongoing debate as to what should be considered the object of a learner's activity [70]. Many applications of AT in the context of education position "the learner's knowledge" as the object of learner's activity, and claim that enhancing the learner's knowledge is the goal of the activity with higher knowledge as the outcome [163,170,177]. This view is criticized as one that assumes a perspective detached from the actual motives of the learner, which are likely to be related to the current situation and the social context: better grades, finishing the course, securing a position within the class or in society, etc. [174]. Others see the school text and other materials as object, while also observing that in traditional school settings, this object is detached from motives and the intended outcome [70]. Jonassen [98] has already pointed towards a tendency to establish overly abstract objects in AT studies and calls for a deeper yet pragmatic reflection on this component. Since the activity is constructed from the subject's perspective [103,159], we think it necessary to reconsider what the object of learning is. We claim that the learner is oriented towards the learning task which she strives to complete. This might be preparing for an exam to complete the course and acquire relevant qualifications, solving a homework assignment to collect bonus points, or preparing a presentation to establish expertise among co-learners or in the classroom. Through the activity the learner transforms the learning task from unsolved to solved or complete, meanwhile identifying her knowledge gaps or insufficient skills, which we refer to as *competency gaps*, and filling those gaps through acquisition of capabilities by consulting the school text, teachers, co-learners, and other sources. In this way, she obtains a *new competency*: that of solving the learning task. This does not however, mean that the learner's competency prior to the activity is irrelevant; the notion of a learner's competency as an object might be relevant to the teacher [70], who strives to make progress with regard to the curriculum. The teacher uses learning tasks to control the learners' knowledge and direct their efforts to discover competency gaps and fill them [119,146]. Accordingly, the intermediate collective objects are the gaps in learners' knowledge and skills, and the ultimate outcome is the completed learning task and the competency to solve this task, which, for the teacher, implies that the learner made a validated curricular progress, and for the learner implies success in obtaining the grade, passing the exam, or receiving a formal or informal acknowledgement of her qualifications.

**Activity Theory in Information Systems**

In the last three decades, AT was adopted in information systems (IS) and related computer-science sub-disciplines like HCI or computer-supported cooperative work (CSCW). Increasing application of AT in IS aligns with the quest for a powerful meta-theory that accommodates the contextual and situational nature of IT as an alternative to earlier decontextualized rationality-based theories [13]. First, it was used to analyze IS interventions and resulting organizational change, for instance, in healthcare [3] or in web-based communication [199]. Second, it was applied to inform the design and development of digital tools while moving the focus from a single-user interaction to activity in a social and material context [12,29,99]. Finally, it was proposed as a core theory for practice-oriented research in HCI [102,114] – the focus of which has moved away from labs and experiments to real-world fieldwork and applications. Despite the sustainable impact on some sub-communities, like the European CSCW, the use of AT for analysis of the most recent and innovative developments in technology remains scarce.

This gap might result from the unclear position of technology in the activity system. While most IS or HCI researchers identify digital technologies with tools [3,102,114], opposing voices [159,160,174] claim that current technological developments go beyond the notion of a tool as defined in status-quo AT. Referring to social media and Web 2.0, Rückriem [160] doubts whether the current generation of AT can accommodate those technological developments and proposes to include the notion of a *medium* within the AT model. Leontyev concludes that the computer technologies form a new leading medium – not a material one but a digital one – and that tools exist within a reality shaped by the new medium [160]. This allows for a more contemporary re-conceptualization of tools as singular functionalities in a complex sociotechnical system.

Another critique of AT is related to notions of subject and human agency. The core of an activity is an object-oriented, motive-driven act of a conscious subject [102,114]. Consequently, being a subject is tightly related to being a human, an animal, or a social entity (e.g., an organization), because only those can do things driven by an inner biological or cultural need [102]. Non-biological entities manifest intentions of others, so are capable only of conditional and delegated agency [102] – they engage in lower level actions and operations [98,102], but not in meaningful activities. However, recent developments in hybrid intelligence [43], in which humans and machines establish highly interwoven assemblages with agency, challenge those assumptions [11]. In such assemblages, engagement in an activity can emerge through shared cognition between a human and a machine. Moreover, following Leontyev, Rückriem claims that digital technologies enable collectivization and externalization of human organs, including human mental abilities, such that computers could conduct actions detached from humans [11,160]. Consequently, it might be possible to acknowledge digital technology not only as a medium or tool, but also as a subject or member of the community. Even though conscious artificial agents are futuristic, we are surrounded by agents which *seem* to possess motives or needs and act accordingly. Rozendaal et al. [159] use Dennett's theory of intentionality [46] to propose the notion of *objects with intent*. Things like a lamp, a clock, a door or a coffee machine have the ability to act autonomously. Dennett claims that because humans attribute intentions to objects, their interaction with the world is altered: based on the ascribed intentions, humans can explain behaviors of animals and non-living things [46]. Applied to computers and digital agents, the theory of intentionality aligns with the *computers are social actors* (CASA) paradigm in which humans apply social heuristics, rules, and expectations to interaction with computers seen as having social attributes [138,139,157]. An ongoing scientific debate concerns the claims and implications of CASA, delivering both empirical studies supporting the theory and others questioning its assumptions [5,66,74,117,135,176]. On the topic of motives and needs, humans have been shown to ascribe intentions to robots and avatars [128,143,164] as well as algorithms that directly impact their actions [176]. However, the existing results are not conclusive on what exactly causes a human to assume that a DA acts according to an internal, autonomous intention or one infused by the designer [128,164]. For example, Rozendaal et al. [159] conclude that a tool, such as a ball mediating a game, can be perceived by the human subject as a quasi-subject one moment, and soon after as a tool without agency. Overall, the idea of agency and its centrality to the definition of subject in an activity system is a challenge given the rapid development of new autonomous systems.

## *Learning-with-PA Model (LPAM) – Activity Theory for Pedagogical Agents*

Despite some shortcomings and limitations, activity theory offers a framework with an extended history in pedagogy and in IS, and previously has been applied successfully to analyze and classify the impact of technologies like serious gaming, mobile learning, or e-learning in educational contexts [25,31,70,81,104,124]. It was however, never applied to study of PA. Still, we argue that AT is useful for a conceptualization of *learning with pedagogical agent* for several reasons: 1) The rationale of the theory is to understand human development and learning and so it has been used extensively in pedagogy, and its components are established in this context, 2) AT acknowledges the social and material character of learning as opposed to cognitive or mental processes of the learner – so AT seems well suited to an understanding of the impact of a PA, and 3) AT can be used for analysis as well as for designing interventions – by proposing this framework, we hope to contribute not only towards the understanding of PA but also to designing such agents. Yet we acknowledge that deriving a framework based on AT is a challenge and requires significant conceptual adaptations. So our model represents an ensemble of interdependent activity systems associated with the activity *learning with a pedagogical agent*. We refer to this model as LPAM (*learning with a pedagogical agent model*) and to its two activity systems as LPAM-LAS (*learning with a pedagogical agent model – learner's activity system*) and LPAM-PAAS (*learning with a pedagogical agent model – pedagogical agent's activity system*). The LPAM and its activity systems are presented in Figure 6.

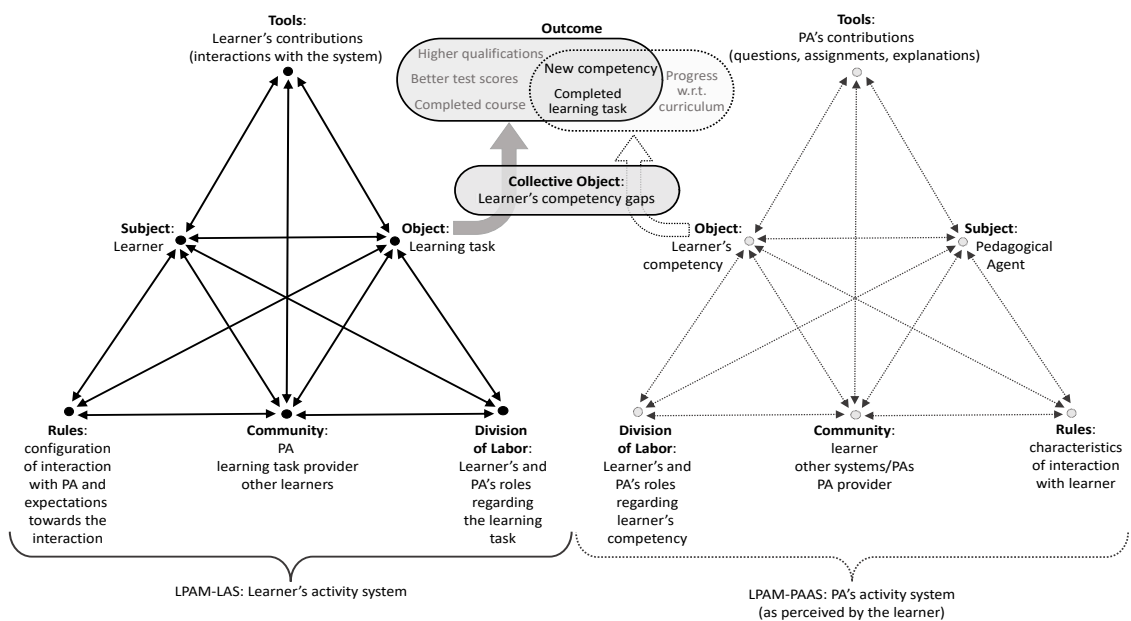

*Figure 2. LPAM: Model of interdependent activity systems involved in learning with a PA based on Engeström's conception of interdependent activity systems [68,70,72] and Dennett's theory of intentionality [46]*

We rely on Engeström's [68] formulation of an activity system as consisting of the six interconnected elements illustrated in Figure 5b. We also use Engeström's concept of the interdependence of activity systems (cf. Figure 5c). Interdependent activity systems consist of at least two activity systems which have a partially or fully shared outcome [72]. Typical examples are provider-customer relationships, hospital-caretaker-patient relationships, or teaching-and-school-going [70,72]. In interconnected activity systems, objects move from the situational "raw material" state through a collectively meaningful state to a set of transformed objects, i.e., *outcome*, including the shared outcome [71]. In LPAM, this embraces the transition from the *learning task* and the *learner's competency* over collectively identified *competency gaps* which prevent the student from completing the task, to the *completed learning task* and *new competency*. The *completed learning task* is a transformation of the *learning task* because it changes the ontological status of the task, e.g., if the learning task is a homework assignment missing a solution, the completed learning task is the assignment *with* a solution; if the learning task is the content that needs to be learned for an exam – the completed learning task is the memorized content. It is also a transformation of the *learner's competency*, which is expanded by the activity if the learner already possessed relevant competency or established when solving the learning task if she did not. Accordingly, the *new competence* is also a transformation of the learner's competency and the task: through the learning task, the learner together with the PA, identifies the missing competency and attempts to acquire it to solve the learning task. In short, the new competency is the learner's ability to solve the given learning task. In formulating additional outcomes for each side (e.g., better test scores, progress through the curriculum), we allow for the fact that participants often have pragmatic motives when engaged in a learning activity [98,174]. Importantly, interdependent activity systems not only share objects and/or outcomes, but might transfer resources or rely on the same context, as shown by examples from previous AT research in pedagogical contexts [70,71,72].

In the proposed framework, we also employ the theory of *intentionality* by Dennett [46] going beyond what Rozendaal et al. [159] proposed about the agency of digital agents. LPAM-PAAS describes an activity system with the PA as a subject. This clearly challenges the need-based agency assumption of AT, because existing PAs are far from being conscious entities with their own needs. However, LPAM-PAAS is not a system found in the external world, but rather an activity system that might be assumed or perceived by the learner who ascribes intentions and agency to the PA, as is likely if the theory of intentionality is true. We argue that by attributing intentions and agency to a non-living entity, one also establishes a hypothetical activity system that explains the actions of this entity and its development (e.g., changing intentions or actions). Earlier research on perceptions of intentions in artificial agents found the process to be complex and may rely on human perceptions of artificial agents in general, the task at hand, as well as the context [128,143,159,164]. Accordingly, as explained by Dennett [46], humans do not simply ascribe intentions to non-living objects, but construct them based on social experiences and resemblance to known situations. So, it is likely that the learner will use *learning with a human* as a frame of reference and make an assumption based on that. We claim that LPAM resembles an interaction between a learner and a member of teaching personnel, e.g., a teacher, a tutor, or a teaching assistant [70,71,174,175]. Owing to this construction, PA does not reside in any other component of the AT – instead we can focus on its actions, modes of interacting, role, tasks, etc. and analyze which of those aspects impact the outcome to what extent.

In summary, we propose to complement the need-based agency assumption of AT with *perceived agency assumption* which implies that subjects are not only engaged in and experience their own activities but are able to construct perceptions of other beings' activities whenever they assume an entity to follow intention and exhibit agency. What emerges is a *perceived activity system* described from the perspective of the perceiving subject and not the subject engaged in the focal activity. This addition aligns with earlier observations that subjects consider activities of other subjects when engaging in their own activities as observed by Engeström [70]. Thus, we claim that the extension is not at odd with the foundations of AT and bears potential to explain how interdependent activity systems function. We next describe the components of LPAM and relate it to the AT.

LPAM: Learner's activity system

**Subject**: *Subject* is the active entity in an activity system. It engages in an activity to fulfill its needs and undergoes changes by engaging in the activity [99,102]. It was natural to see the learner as a subject in the given activity – she is the one undertaking effort to acquire new skills or memorize new content. In the subsequent literature analysis, this AT component is used to describe various categories of learners (e.g., children, high-school students, university students, professionals).

**Object**: *Object* is the focus of attention during the activity and the concern of the subject [71] – it meets the subject's need [102]; the transformation of the object provides a way to meet that need; the transformed object is the ultimate product of an activity system [98]. We claim that the activity of the learner is oriented towards a *learning task*, which undergoes a transformation when being worked on. In the subsequent literature analysis, we will differentiate among various types of tasks supported by PA (e.g., solving equations vs. learning to write).

**Tools**: *Tools* encompass artefacts, sign systems, theories, procedures, techniques, ways of interaction or even 'automatized actions' [98,102,160] used to mediate between the object and the subject. There are manifold tools involved in learning new content or mastering new skills (textbooks, graphics, explanatory videos, conversations with teachers or co-learners, demonstrations). During *learning with a PA* (LPA), the student engages with the learning task by using her materials (school text, paper, pen, word processor). But of primary interest to the current study is the fact that she also uses interaction with the system to solve parts of the learning task, to specify and extract subtasks, or to express competency gaps. In the following analysis, we explore what contributions are allowed by PA and how PA uses the learner's contributions to adjust the activity.

**Community**: Activities are contextually and socially bound, individuals within a *community* support various activities so that the subject and the community form an assemblage of agents oriented at a common object [11,98,99]; community can also be a source of motives and needs [102]. During LPA, the agent provides the intermediate context and support for the activity and fulfills the role of a social member of the community. But we can extend the context to consider the person who provides the learning task (e.g., teacher) or other learners (if any are involved). In the subsequent analysis, we will focus on the characteristics of PA which make it a member of a community, like elements of the design that make it appear intelligent and conscious, interactions with or references to other learners, teachers, current societal context, etc.

**Rules**: *Rules* are means for mediation between the community and the subject [102]. They define what behaviors are appropriate and guide interactions within the community [98,103]. Rules include explicit regulations as well as implicit conventions, social norms, or relationships [98,99]. In the above activity system, rules regulate the interactions between the learner and the community, primarily PA: they specify how the interaction with PA is possible and what related expectations the learner can have. Next, we will analyze how various configurations of interaction (e.g., what actions are allowed from the learner) impact the learner's outcome.

**Division of Labor**: The *division of labor* mediates between the community (including the subject) and the object of activity [102]. It refers to horizontal distribution of tasks as well as to the vertical distribution of power and status [68]. In the following analysis, we focus on the distribution of responsibilities that mediate among the learning task, the PA (as a member of the community), and the learner. Both PA and learner are members of the community. PA can play various *roles*, e.g., to guide through a task, to check for completeness, to explain the task, to help identify information necessary to solve the task. However, the learner also can take on different roles: a tutee vs. a co-learner.

LPAM: PA's activity system

**Subject**: The pedagogical agent is a *subject* in the *perceived activity system*. It is the entity that can be ascribed agency and exhibits specific needs or motives. We will explore how the PA expresses its agency and whether this has an impact on the learning outcome.

**Object**: Pedagogy is seen as "any conscious activity by one person designed to enhance learning in another" [186:17]. Consequently, the PA activity is oriented at the learner's competency that should be transformed in the activity – this positions the learner's competency as the *object* of PA's activity. We will consider various types of competencies as characterizing this component.

**Tools**: The basic way to mediate between learner's knowledge and the PA is interaction with the learner. Only based on an interaction can PA assess and impact the competency of the learner. Thus, interactions with the learner have a *tool* character for the PA. The agent can employ various techniques or artefacts, which we denote as *contributions*, to help the learner identify competency gaps by asking specific questions or providing assignments, provide instructions on how to address the gaps, or provide material and explanations that contribute to closing them. In the subsequent analysis, we will identify whether and how the choice of technique or material impacts the learning outcome.

**Community**: Apart from the learner with whom the PA interacts, it has its own context of technological connections including the system it is integrated in or other PAs and their organizational context, including the person or institution who built or provided the PA to the learner. The latter might have essential impacts on how the learner views the PA. In our analysis, we consider integration of the PA in software and hardware systems (e.g., part of a larger intelligent tutoring system or a standalone PA), its origin (e.g., developed by the teacher vs. an external entity), and "collaboration" with other PAs.

**Rules**: In its interaction with the community and the learner, the PA can fulfill various visions of how a PA should act, reflecting the social rules and norms in traditional learning environments. We will consider general adaptations concerning the style of interaction, which may trigger specific expectations from the learner, as characteristics of this component.

**Division of Labor**: The PA and the learner can have various roles and functions regarding the learner's competency. For instance, a PA can act as an examiner or induce reflection processes in the learner. The subsequent analysis will consider the impact of such roles on the learning effect.

## METHOD

To understand how the design of a pedagogical agent impacts the learning outcome, i.e., the amount and quality of acquired knowledge or mastered skills, we analyze previous studies addressing the topic of LPA. Specifically, we consider studies in which learning outcome was measured as a dependent variable and at least one aspect of the pedagogical agent was manipulated (e.g., male vs female voice), or a quality of the agent (e.g., voice quality) was treated as an independent variable (also, explanatory variable or input variable). In doing so, we identify aspects of a PA with potential influence on the measured learning outcome. Of course, by following this method, we cannot measure or consider any potential interferences among the various variables, though it is also impossible that these aspects are independent from each other (e.g., projecting humanoid character is hard if the learner can only hear but not see an agent). Also, the variety of measures, manipulations, and experimental designs applied in the considered studies makes it nearly impossible to treat them as equivalent in any respect, except that they focus on an individual learner interacting with a PA. Therefore, we resort to identifying and providing an overview of the relevant aspects and classify them according to the framework described in section 2.3 and illustrated in Figure 6. While this cannot provide a conclusive answer on whether an agent designed to combine the various features will have positive impact on the learning outcome, it is sufficient to establish a catalogue of design decisions and identify areas that require further investigation.

To select the relevant studies, we employed a systematic, concept-centric literature review [17,121,187] using the following procedure. First, we selected the sources to be included in the initial search. We decided to consider the top 25% of outlets from the following disciplines according to the Scimago Journal Ranking (SJR) as used in the Elsevier Scopus database: Education, Educational Psychology, Computer Science, Artificial Intelligence, and Information Systems. This resulted in a collection of 1,137 outlets including journals and proceedings of scientific conferences. Since Scopus provides search capabilities for title, abstract, keywords, and other bibliographical data for all those outlets, we limited our search to this database. Second, based on seed literature and recent publications from the most relevant outlets in each discipline (e.g., *Journal of Computer Assisted Learning*, *Computer and Education*, *International Journal of Artificial Intelligence in Education*, *International Conference on Human Factors in Computing Systems*) we identified a set of terms used to denote a PA (or a related concept) and a set of the most influential authors, whose publications were used to adjust the list of terms. The list included 31 different terms (e.g., *natural language tutor*, *animated pedagogical agent*, or *tutoring system*). We reduced the list to the following query for search in Scopus:

```
("conversational"|"dialogue-based"|"dialogue"|"natural-language"
|"intelligent"|"animated"|"embodied"|"social"|"cognitive"|"adaptive")
AND ("pedagogic* agent"|"pedagogic* robot"|"tutor* system"
|"tutor* agent"|"tutor* robot"|"robot tutor"|"computer tutor")
AND ("educat*"|"learn*"|"stud*"|"teach*").
```

Importantly, the operator 'AND' does not imply that the terms are concatenated, but it means that the terms on the left and on the right of the operator need to appear in the title, abstract, or the keywords, but may be separated by other words. '|' stands for logical disjunction. The search in Scopus resulted in 775 articles between 1973 and August 31, 2020, with 431 published in 2010 or later. Third, we excluded articles according to the following criteria applied in the given order: (a) conceptual articles without empirical basis (n=60), (b) meta-studies and reviews (n=18), (c) studies addressing ITS or other tutoring systems that do not allow for natural-language interaction with the agent (e.g., systems that control user input during a task to later mark the step in which the learner made a mistake; n=607), (d) studies which do not measure learning outcome or learning performance in any form (n=82), (e) studies conducted in contexts where learning outcome is impossible to assess (e.g., creative teams; n=13), and (f) studies in which a group of learners interact synchronously with a PA (e.g., PA as a classroom assistant; n=5). After application of the exclusion criteria, we selected 51 articles for further analysis. Based on the backward and forward search rooting of those articles and the partially relevant literature reviews [9,96,113,141,148,194], we identified 74 more potentially relevant articles which, after application of the exclusion criteria, expanded the number of selected articles by 13. They passed the same filtering procedure as the keyword-search-based results. Overall, 852 were considered during the selection process, and 64 were chosen for the analysis. Application of the criteria was conducted by a full-time researcher and all controversial cases were discussed with other authors, so the final in-or-out decision was made collectively. Lastly, a second postdoctoral researcher screened all selected articles for their agreement with the criteria. All 64 selected articles were included in the subsequent analysis.

The analysis was oriented at identification of contextual and design aspects that influence the learning outcome during LPA. We proceeded as follows: All relevant studies were described in terms of the measured learning outcome (dependent variables), differences between treatments, i.e., manipulated aspects, explicit control variables (independent variables, or factors), and the relationship between the independent and dependent variable. This enabled us to identify those aspects of the design or context that were reportedly tested concerning their influence on the learning outcome [2]. If an article describes more than one independent experiment, each was considered separately. If an experiment uses factorial design (two or more factors define experimental condition), each factor was treated separately as a design aspect and explicit results for that factor were considered. If such results were not provided and no direct impact of the factor on the learning outcome could be established, but the results indicated that the impact is moderated or dependent on another variable, the interaction between the factor and the dependent variable was marked as *mixed*. Similarly, if the experiment provided contradictory results (e.g., highest level of factor A leads to high learning gain, medium level of factor A leads to negative learning gain, and low level of factor A leads to high learning gain again), we marked them as mixed. Otherwise, we classified the interaction as univocal and described the interaction as positive (learning outcome grows if factor A grows), negative (learning outcome falls if factor A grows), or having an impact in explicit terms (e.g., condition-1 > condition-2 means that learners in condition-1 have higher learning outcomes than in condition-2). If no significant impact could be identified, the interaction was marked as *neutral*. The identified factors were matched subsequently with the LPAM components (cf. Figure 6) in a workshop involving all authors of the current manuscript and guest researchers. This should guarantee that matching between the theoretical categories of the AT and the design or context aspects considered in the studies is reasonable for a broader public with varying degrees of experience in AT and/or learning technologies. Further, each study was described in terms of the learner category and PA category and characteristics. Those descriptions were used to characterize the *subject* in both activity systems depicted in Figure 6. The next section summarizes the results of a systematic review.

---

[2] The underlying studies most frequently use causal language to describe the relation between an aspect of the LPA and the learning outcome. Those studies either do not directly discuss causality or they infer causal relationship from education theories and from temporal order of intervention and measurement. However, few examples are cautious enough to point out that their results describe a correlation [22,78,91,149], but even those studies make causal claims based on the results. In this study, we consider studies independently of their approach to causality and follow them by assuming that changes in the LPA *cause* the observed transformation of the object. This view aligns with AT.

## RESULTS

### Learner and PA as activity subjects

The examined articles contain research on several different activity subject categories, both on LPAM-PAAS and LPAM-LAS. With respect to the learners, the subjects are differentiated by age and then their educational context, as described in

Table 5. Primary school children's ages range from 5 to 11 years, while secondary school students range from about 12 to 15 years. A dominant tendency towards PA research in higher education, which includes college and graduate students, is clearly discernible. The category of adults includes participants of all ages, independent of their educational status.

*Table 1. Number of articles by learner subject category*

| Learner Subject Category | Primary school | Secondary school | Higher education | Adults |
|---|---|---|---|---|
| **Number of projects** | 7 | 7 | 46 | 4 |

In terms of PA subjects, four categories can be distinguished: chatbots, cartoon agents, human-like agents, and robots (cf. Table 6). The category of chatbots includes pedagogical agents that solely use text-based or voice-based natural language to communicate and do not have an embodied representation of any kind. For instance the NLtoFOL system, developed to train learners to translate natural language to first-order logic, has been equipped with a chat-based agent to guide learners through a series of tasks by giving them hints and feedback via automatically generated natural language sentences [152]. Cartoon agents depict an animal or an abstract being that does not resemble humans but might have human-like mimic or gesturing behavior. They use either text or speech to communicate with learners and can be either animated, including anthropomorphic features, or static, with a trend toward animated agents. Such agents are used for a variety of tasks, including examination of the effect of facial expressions on learners [2], whether the simple presence of a PA [198] in different forms [24] (no agent, voice only, static agent, full agent) makes a difference in the motivational and learning outcomes, or to test how a gesturing agent performs in comparison to a non-gesturing agent [34]. In contrast to cartoon agents, human-like agents possess visual features. Although computer animated, they are purposefully designed to look human. Some of them are simply talking heads, others show a virtual upper body or even full human body. While some talking heads are explicitly used to test the effect of PA enthusiasm [125], personality [10], and rapport behavior [112] on the learners, others are again used to test how the (visual) presence or absence of a PA affects learning outcomes [131]. Full-body human-like PAs can be used to combine instructions, practice and feedback in one agent by having the learners perform actions on the body and the environment of the agent [23]. Lastly, robots are used least frequently and predominantly with young children as learners. Using speech and physical signaling, these robots interact with children and teach them multiplication [110] or play word-learning games [28].

*Table 2. Number of articles by PA subject category. If more than one category was tested, we count the paper as belonging to the more advanced category*

| PA Subject Category | Chatbot | Cartoon agent (e.g., avatar) | Human-like agent (e.g., virtual human) | Social Robot |
|---|---|---|---|---|
| Number of projects | 10 | 11 | 35 | 8 |

**LPAM-based Analysis of Impact on Learning Outcomes**

The articles relied upon consider very diverse subjects. Yet, we found even more variety in relevant outcome measures. In particular, learning outcomes are measured using a broad selection of constructs ranging from multiple-choice verbatim knowledge questions, i.e., questions directly from the PA asking for information [22], to measurement of whether the learner can apply the knowledge to solve a task in mathematics or logics [34,152] or a vocabulary test [28], to deriving measurement, e.g., knowledge monitoring assessment comparing the learner's judgment about her knowledge with task performance to assess her metacognition abilities [105]. In the last case, the learning outcome is the learner's self-assessment competency. Sometimes studies refer to learning gains as tested by a pre-treatment and post-treatment test (short, pre- and post-test), others use only a post-test, mostly administered directly after treatment, but sometimes delayed. All this makes it impossible to combine results from separate studies and provide a conclusive, bottom-line answer on how an aspect impacts the learning outcome, even if some studies attend to the same or a similar aspect. Further difficulty arises from the specific experimental conditions: if a design aspect, e.g., presence or absence of a voice-based narration, is tested with different PA types, e.g., a social robot and a virtual human, and with different learning task, e.g., foreign language conversation or solving mathematical problems, it is hard to argue for the comparability of the studies. Consequently, we provide a catalogue of tested aspects and their impact on learning outcomes in a study-by-study manner in the virtual appendix available from the Journal's website.

Figure 7 summarizes the quantitative analysis and explicitly links it with the LPAM. The figure uses various shadows of red for LPAM-LAS and various shadows of blue for LPAM-PAAS. The analysis shows which components and which aspects were attended to most. For instance, the most studies attend to the subject of LPAM-PAAS (overall 30 studies, of which twelve suggest positive or negative influence of an aspect on the learning outcome, seven provide mixed results, and eleven provide neutral results suggesting no influence on the learning outcome). 16 studies attend to how visual and bodily presence of the PA impacts the learning outcome: six suggest that more visual and bodily presence will enhance the learning outcome, one suggests the opposite, i.e., that reducing the visual and bodily presence will enhance the learning outcome, three studies provide mixed results, and six studies suggest that there is no correlation between the visual and bodily presence of a PA and the learning outcome. On the LPAM-LAS side, most studies (nine of them) attend to the learner, i.e., the subject component in this activity system, of which six investigate the relationship between the prior knowledge of the learner and the outcome resulting from a learning episode with a PA. Four of them yield mixed results, suggesting that there is a relationship between the prior knowledge and the learning outcome when the learner uses a PA, but the relationship is not monotonic. for example, higher or lower prior knowledge does not imply better or worse learning outcome. Instead, the relationship involves moderators or sub-categories.

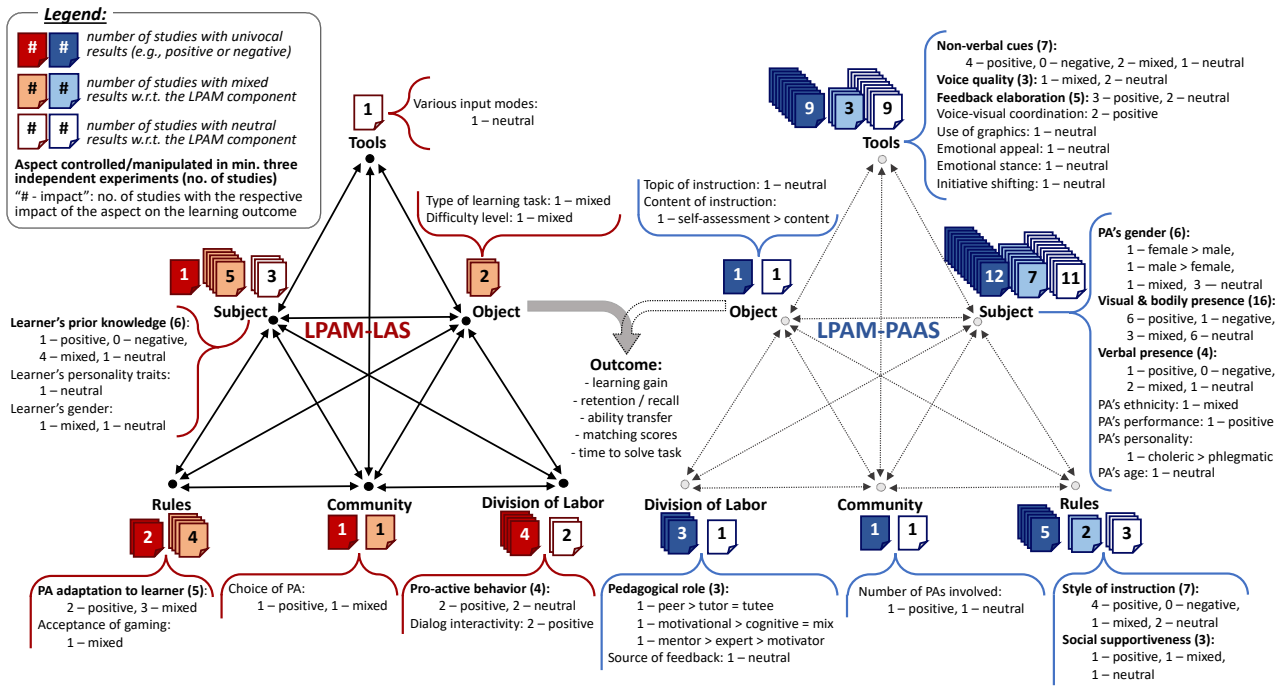

*Figure 3. Results overview of the literature review depicted in the LPAM including relevant design aspects.*

The analysis shows that the available studies attend to many varied aspects of the PA and of the whole activity of LPA. Whereas some factors like the PA's gender or visual or bodily presence were controlled or manipulated by multiple studies, sometimes even without direct reference to each other, others like the PA's age, ethnicity, or learner's input modes are addressed by singular studies only. A first glance at the LPAM including the results of the analysis in Figure 7 confirms that the distribution of the analyzed studies across the components of the model is imbalanced, with more focus on characteristics of the subjects or tools, and consistently less attention to the context and community. Further, we observe that many studies yield mixed results, which suggests that the interdependencies between the various factors are complex. In the following we address the findings of the analysis.

The results indicate that a PA or, in a general sense, LPA is thoroughly a sociotechnical system. Specifically, one can notice the very tight interconnection between the technical and social component: social effects such as adaptation to the learner or a socially supportive stance are accomplished by technological means such as an avatar graphic blended with the learning interface or a specific text-to-speech engine [115,117,121,134]. Further, one can observe the alignment of instrumental objectives, e.g., better test results, and humanistic ones, e.g., ethnic tolerance [146]. In fact, going beyond the learning outcome, many articles measure emotional engagement, satisfaction, or willingness to collaborate [e.g., 32,94], values considered social or humanistic [18,177]. In short, LPA is a typical sociotechnical interaction, not because we present it as such, but because the insights of the underlying studies confirm a mutual dependency between technical and social aspects, as well as instrumental and humanistic outcomes. Moreover, this dependency is visible in the interaction between a PA and a single learner. However, the studies rarely view a detached learning episode disconnected from classes, though a few examples of artificial learning content exist [22,34,42,43]. Most studies are conducted in areas relevant to the learner as a member of an organization, sometimes within coursework, such that PAs are administered as support for homework, an individual learning episode during a class, or to complete other curricular duties [9,117,118,141,176]. So the activity is embedded in a broader organizational context, likely generating effects beyond the interaction between the learner and the PA. However, those effects are unattended and call for more attention to the larger context of the learning activity.

## DISCUSSION

Applying the AT perspective to the studies on PA shows that much effort is put toward understanding how inherent attributes of a PA or the way it interacts with the learner influence the learning outcome. The studies consider various combinations of visual, bodily, and verbal presence, as well as implement various behavior schemas employed by PA. However, many areas remain barely touched, especially the collaborative aspects of the activity: rules, division of labor, and role of the community attract less attention from researchers despite a large variety of settings to be examined. Even though PA have attracted some interest from IS [191,192,194], attending to the collaborative components of the learning activity could position this class of systems at the core of IS interest [13,15,162]. The sociotechnical nature of LPA and the likelihood of its becoming increasingly pervasive with natural language processing capability and increased distributed learning created by the COVID-19 pandemic, calls for IS researchers to engage. In Section 5.1, we approach the subsystems of the LPA, identify emerging questions, and link them to IS discourses. We claim that research on improving learning by means of PA can benefit from IS competence as presented below. However, IS can benefit from PA research to inform the design of DA beyond education as well. We attend to this aspect in 5.1 and explicitly discuss it in 5.2.

## Critical Reflection on the State-of-the-Art

LPAM: PA's Activity System

**LPAM-PAAS – Production Subsystem:** This subsystem focuses on how the subject transforms the object by means of the available tools. Within the LPAM, this is the subsystem with the most contributions, primarily due to the many studies of the tools and the PA as a subject of the activity. The studies of the production subsystem try to establish links between technical and methodological aspects of PA's design and the learning outcomes. Most of them manipulate the PA, so that in one condition it is a little bit more human-like than in the others, e.g., presented by an avatar vs. not, or using richer channels for transmitting the message, e.g., by combining voice narration with gesturing vs. no gesturing. Accordingly, one learns how altering single technical aspects in the design of the PA can generate significant changes in outcomes. The results suggest that more is better: more anthropomorphic features, more parallel channels and more presence leads to better learning outcomes, with only single studies that contradict this claim [2,37,38]. This sounds indeed like *production* – the more resources provided, the higher the performance. However, only two studies attend to the characteristics of the object. This leaves open whether and to what extent insights about the impact of design aspects can be transferred across types of instruction content or bridge diverse competency gaps identified in the learner. In short, it is hard to assess the generalizability of the results – the large majority of the considered studies employ PA in a single-topic scenario and in a very specific context, with few exceptions [7,8,109]. On the other hand, a body of knowledge emerges on a few specific features, like gesturing vs. no gesturing, if one considers the entire history of discourse around this topic. IS research, especially the design science research [150], has reported frequently on difficulties related to the transfer of design insights and calls for additional checks on the transferability or reusability of the results [1,92,153]. Reusability of design principles and reproducibility of intervention results is a topic of growing relevance to the IS community as confirmed by recent publications [18,92,93]. We claim that there are significant opportunities for mutual benefits for the IS domain and PA discourse: many PA studies attend to design aspects in an unsystematic manner, without framing the experiments as reproduction attempts – IS and its long research tradition on development, evaluation, and reusability of design principles offer a methodological underpinning for such studies and can enhance their validity [93,94,150,180]. Yet PA discourse re-attends to specific design aspects again and again while technologies or learning contexts change, so it is an interesting case for studying the emergence and re-evaluation of design principles since the 1990s [113,141]. IS research can build upon past studies of the *production* of competencies through use of PAs' resources by formulation and evaluation of generalizable design principles, proposition of new, methodologically-grounded evaluation strategies for LPA that go beyond single experiments, and by establishing a coherent framework for the development and evaluation of PAs inspired by almost 20 years of design science research (DSR). At the same time, it bears potential for the update of the DSR framework itself: while DSR focuses on design artefacts, we suggest that the object of DSR can be experiences rather than artefacts. As suggested by some of the PA studies that identify complex interdependencies, it is the whole situation and experience of the subject that produces learning outcomes. Especially when designing a DA, one designs not an artefact but an encounter that is likely to be perceived as social to some extent. The notion of a design artefact might be misleading in this context, so we see potential for redefining it towards a more inclusive notion.

**LPAM-PAAS – Distribution Subsystem:** The distribution subsystem links the object of the activity with the community through responsibilities and roles. This subsystem has the lowest coverage in the considered studies across the whole LPAM (cf. Figure 7) and the relevant discourse has focused on the pedagogical role of the PA(s). Specifically, we did not identify any study that tries to establish a link between the distribution of responsibilities in the broader organizational context of learning and the learning outcomes. The scarce coverage of this topic invites contributions from the IS discipline. We identify the following opportunities: First, past IS research on e-learning has shown network-based reasoning is useful to understanding how e-learning is appropriated and what roles emerge [35,79]. From the perspective of the learner, PA – even when interacting with the learner alone – might seem involved in the social network of teachers, tutors, and school personnel, and therefore those individuals might influence the learner's experiences with the PA and, consequently, the learning outcome. So, it is necessary to develop ways to integrate the PA into the organizational context of learning and raise questions about how it interferes with various teaching formats (e.g., blended learning vs. traditional classes), organizational structures (e.g., research university vs. distance education), and how it interacts with various actors (e.g., teacher, school, department head). The latter aspect of the network points to a second interesting research opportunity. Given that the PA will get increasingly intelligent because of AI development, new configurations of work can emerge and revolutionize the organizational context of education altogether. IS has recently started to explore *the future of work* and the distribution of competencies between humans and AI [16,43,57,58,77,127,167], but has so far focused on the business context. We claim that identifying the optimal distribution of tasks between human and digital teachers will become a very important topic in education for the next decades, and IS can contribute by leveraging its concepts and experience with human-AI collaboration. Finally, the analyzed studies did not address the question of who develops and distributes the PA to students. Learning outcomes likely will differ if the PA uses content provided by the teacher, aligned with classroom activities, as opposed to an independent PA, but this claim has not been addressed. Many of the analyzed studies use complex systems developed in a lengthy process due to the shortcomings of earlier technology. However, now anyone can develop a dialog-based agent using free online tools. Recent IS research has leveraged this opportunity by providing even more accessible tools to teachers [190,191,192,195] to help them design their own PAs, changing the distribution of responsibilities in designing and developing a PA. Education is one of the most progressive fields for the design and development of DA by stakeholders who are not IT professionals. Understanding how this rapid empowerment of teachers will change learning outcomes will fuel predictions about the future of work in education. Given that the technical frameworks will become increasingly accessible and easier to use, potentially everyone will be able to create their DA. This has implications for distribution and character of work in the whole of IS [16] – stakeholders might become capable of delegating part of their work involving contact with other humans to machines that they can program themselves. This change might indeed come faster than potential changes implied by emergence of general artificial intelligence. While the community considers enabling factors for complex collaboration between DA and humans [167], one key factor, the ability to develop a DA with simple means, might change the character of work delegation in the near future.

**LPAM-PAAS – Exchange Subsystem:** The exchange subsystem focuses on explicit and implicit norms of the exchange among the subject, PA, and other members of the community. The various aspects of how PA presents itself to the learner dominate the discourse related to the exchange subsystem in LPAM-PAAS [10,107,110]. Evidence from the analysis suggests that informal, casual, and positive interaction with a visually or bodily present PA leads to the most successful exchanges and best learning outcomes. However, this involves a rather static idea of interaction and communication style: the large majority of PAs were implemented to behave the same way throughout the interaction, rather than reacting to changes in context – be it changes within the interaction with the learner or changes in external context, such as passage of time [125,130]. It provides a chance to investigate the impact of external context on the exchange with the PA, especially the material, temporal, and technical context including the relationship between PA and external teaching materials or e-learning systems. According to the socio-material perspective adopted by many studies of situated work practice in IS but also in HCI or CSCW [27,55,59,61,63,145,173], the temporal and material structure moderates the effect of using IT. The socio-material perspective studies how language, action, interaction, organizational practice, as well as societal roles and identities are entangled with human bodies, spatial arrangements, physical objects, and technologies [27,145]. None of the selected papers applied the socio-material lens to the interaction with PA and the resulting dependency on the wider context, but there exist convincing accounts of teaching and learning generally as a socio-material undertaking [75,132] including the role of IT in a socio-material context [97]. Given the IS association with socio-materiality, we see an opportunity for research efforts on the impact of material context of interaction with PA on learning outcomes and design efforts to make PAs adaptive to changing circumstances to facilitate the exchange with the learner. In particular, attending to how to introduce an agent in a way that makes it appear natural and well-integrated is an open challenge for design of PA and for the design of DA in general [58,64].

**LPAM-PAAS – Consumption Subsystem:** The consumption subsystem focuses on the subject as a member of the community in relation to the object. This is the least balanced system in the whole LPAM (cf. Figure 7). On the one hand, we have just two studies attending to the object or community components, and on the other, we have 30 studies analyzing aspects of the subject. This is somewhat symptomatic of the whole analysis: whereas much focus is put on the PA, its behavior, presence, capabilities, etc., few studies focus on the context and how it moderates PA's impact on learning outcomes. Going beyond the material and temporal context, and the organizational context, we want to stress the fact that PA functions within a technological context: it mostly "resides" within an e-learning application [52,53,141,179,198] or refers to one when interacting with the learner [122,123]. To fully transfer the insights about the effectiveness of PAs, one needs to understand the interplay between the e-learning application, which forms a quasi-virtual context for the PA, and the PA: Does the learner see them as one entity or two separate entities? Does this perception affect learning outcomes? How can we transfer insights about the effectiveness of on-screen avatars to social robots or vice versa? Another interesting aspect might be the presence of more than one agent, as in two of the selected studies [52,198], raising some ontological issues. Even though the PAs in such setting are presented (and may be perceived) as separate entities, they depend on each other not only via division of labor, but they run in a single instance of an application and presumably share the same data, even if their presentation is different. We need to conceptualize the interaction between intelligent agents appearing across domains. A promising approach is the multi-agent-system perspective [76,161], which aims at modelling interdependencies between intelligent entities. This paradigm was broadly applied in IS research on automated negotiations, e.g., in power grids or data markets [111,137], but also complex processes like open-source consortia [73] or space mission operations [171]. This flexibility of the multi-agent perspective can be very helpful when analyzing complex ensembles of multiple PAs, teachers, and learners, to understand the dependencies among them, consumption of each other's resources, and the outcome.

LPAM-LAS: Learner's Activity System

**LPAM-LAS – Production Subsystem:** Now we attend to how the learner, as the subject in LPAM-LAS, uses available tools to transform the learning task from *unsolved* into *completed*. When comparing the production subsystems of LPAM-LAS and LPAM-PAAS, one difference is particularly striking – whereas many studies consider what techniques, methods, and media PA uses to communicate to the learner, only one study investigates *how* various input modes from the learner to the PA impact the learning outcome [54]. In other words, we have very limited insight on whether providing the learner with various media to talk to the PA helps or hinders the learning task at hand. There are just two studies trying to link the characteristics of the learning task to the learning outcome [134,147]. This is symptomatic for the analyzed studies and their fuzzy claims about the transferability of results: according to media richness [40,48,101] or media synchronicity [47] theories, it is likely that the preferred medium would be chosen based on the learning task. However, in LPA the choice of medium is limited, so efforts should be made to find the optimal match to facilitate better outcomes. However, many of the selected studies make the case for general tendencies as to what works better and why. Given that media choice theories have a strong tradition within IS, we call for a careful investigation of the media-task fit in the context of LPA, for instance, to identify media-task combinations that lead to optimal learning outcomes. In fact, we observe potential for a fit theory that applies beyond the context of education, and IS is well-suited to propose such a theory.

**LPAM-LAS – Distribution Subsystem:** The focus in here is on the distribution of responsibilities and roles among the learner, PA, and the learner's context. The distribution subsystem in LPAM-LAS is dominated by the aspect of pro-activity of the PA in interactions with the learner. Specifically, this refers to the question of distribution of control and rights in the dialog, who can control when and what contributions are made, along with the distribution of responsibilities, e.g., who is responsible for starting the learning episode and keeping it going. The results suggest that a PA that uses prompts and takes initiative might positively impact the learning outcome, while also motivating the learner to learn [65,87,129]. This aligns with the general potential of technology to be capable of persuasion and nudging. The persuasive technologies discourse has emerged in the last few years frequently in relation to bringing about change in a conscious and self-determined human, who engages with the technology to improve her own motivation and ability to pursue the change [60,62,84,144]. Digital nudging developed in parallel as a way to provide subconscious or semi-conscious behavioral triggers to guide the user in the desired direction [49,188]. The concepts of nudging and persuasion come originally from the study of human conversation [32,56], but popularization of online commerce, wearables, self-improvement mobile apps, etc., has moved the focus towards digital persuasion and nudging. In interaction with conversational agents, these two lines of research merge: on the one hand, we have natural-language conversation which can be constructed in a persuasive manner, and on the other hand we have a digital agent, which may employ more sublime means of persuasion. As the results of the review show, just exhibiting a more pro-active behavior can have a motivational effect and lead to higher learning outcomes. However, this can lead also to ineffective distribution of responsibility, in which the learner will feel motivated only when interacting with the PA, like people who report less motivation to exercise without their wearable to track themselves [80]. We encourage research to investigate the intended and unintended effects of distribution of roles, tasks, and rights between the learner and the PA.

**LPAM-LAS – Exchange Subsystem:** This refers to the norms governing the exchange between the PA and the learner, as well as the social context of the learner. The exchange subsystem is the one with the strongest coverage in the selected studies. The studies considered in this subsystem address primarily the topic of mutual adjustments happening among the learner, the PA, and the overall activity – adaptations of the system and its components. First, numerous studies address the impact of a learner's inherent characteristics, like her gender or prior knowledge, on the effectiveness of learning with a PA. The results are mixed, but several studies suggest that weak learners can experience disproportional benefits from learning with a PA [30,90,95]. Second, the dynamic adaptation of PA's behavior to the learner's needs or status positively influences the learning outcome [116,122,156]. Accordingly, by contrast to LPAM-PAAS where no adaptation to the context occurred, the exchange in the LPAM-LAS is characterized by a dynamic negotiation of rules between the PA and the learner. This mutual adaptivity elevates the interaction between the learner and PA from information exchange to the level of dialectic interaction, so that a human-computer assemblage emerges, in which the PA controls not only for explicit but also for implicit exchange cues. The need for high adaptability is a theme of the new IS discourse on machines as teammates [74,127,138,167] – only with adaptive behaviors can intelligent machines seamlessly integrate with human work practices. The research on exchange with PA seems to confirm this. Still, the selected studies mostly cover a single cue that the machine controls for and adapts to: boredom [53], uncertainty [78], or performance [156]. However, learning is a complex process and occurs within a complex, dynamic context, such that multidimensional adaptations are necessary. Creating machines capable of complex adaptations is the declared goal of the so-called third wave AI [20]. The results here suggest that highly adaptive PAs will be able to support learners even better in reaching desired outcomes, but this remains to be confirmed.

**LPAM-LAS – Consumption Subsystem:** As in the LPAM-PAAS, the consumption subsystem within LPAM-LAS exhibits some imbalance – whereas there are studies attending to the characteristics of the subject and positioning it within the broader community (e.g., as a weaker or stronger learner), only two studies attend to an aspect relevant to the community: the learner's freedom of choice concerning the PA [134,147]. However, none of the selected studies addresses the broader community of the learner, i.e., other learners and their potential impact on preferences and learning outcomes. Other learners provide the social context for the activity: they might compete with each other, form cliques, or simply influence each other's preferences and strategies towards the PA. Attitude towards IT and social influence as a key factor affecting the acceptance and use of technology has been acknowledged in core IS theories for decades [21,51,181,182]. We hypothesize that only by understanding a broader context of use, including learner's social community, can one explain long-term effects of learning with a PA. We therefore call for field studies observing the use and effect of PA on learning outcomes in the context of learning. Most of the selected studies rely on results obtained in experimental conditions. Even if the experiments were linked to the learner's organizational context, the incentives to perform well in the experiment were detached from normal classes and long-term motives for learning. IS has always stressed the need to understand long-term outcomes of digital interventions and advocated interaction with practitioners when designing and evaluating them [44,108,140,151]. We align with those suggestions and encourage ´effort to conduct long-term studies integrated with learner's goals.

**DA beyond Education**

Apart from the research directions identified above, the theoretical perspective and the results have implications for the application of DA beyond an educational context. We see the current article as a step towards unpacking and understanding interaction between a human and a DA. First, the AT perspective positions the interaction as part of a complex activity involving material, organizational, and social components. The results show that those components, e.g., the relation between the content provided by the PA and that provided in the classroom, as well as the distribution of roles among DA, the learner, and other actors, has significant impact on the learning outcome. We claim that similar effects can be observed in other settings including healthcare or commerce. It is likely that the impact of a DA on a patient's health outcomes will depend on the characteristics of the therapeutic alliance between the patient and the therapist. However, much research on the use and effects of DA was conducted in experimental and exploratory conditions and focuses on the dyadic relationship between the DA and the user. This perspective detaches the DA-human interaction from the social and organizational context. Simultaneously, DAs are claimed to mimic interaction between humans to make the interaction between machine and user more natural. However, interaction between two humans happens within an organizational and social context, is driven by social roles, interactional rights, and organizational identities, and those roles are reflected in material and content used in the human-to-human interaction [55,61,63]. Therefore, it is surprising that the design and evaluation of DA frequently abstracts from it. Even though most PA studies also adapt this detached perspective, several of those selected for the current review explicitly attend to the character of learning as embedded in a broader context. For instance, they explicate the role taken by PA and contrast it with the roles taken by humans in a similar position [100,141,152]. In doing so, they attend to situational scripts, expectations users might have of the agent, how those expectations can be manipulated, or how the agent can be embedded in those expectations. They deliberately treat the design of a PA as a subprocess of designing the whole learning situation, as reflected in the framing by some of those studies: ITS. Even though the definition of an Intelligent Tutoring System often relates to a *computer system*, some describe its architecture in a more general terms [113,152]. In this understanding, ITS embraces more than an intelligent tutor, namely, the technological context, the student model, the teaching model, as well as the teacher, and the learner [4,113]. This article goes further by including other social actors as members of the community, attending to the rules that govern the activity, and acknowledging the division of labor between the actors. This perspective accentuates the sociotechnical character of work and interaction with a DA [162], with implications for design. Creating a DA is not about designing ways of interacting with an artificial agent but involves engineering the whole situation and how it is embedded in the societal and organizational structures. DA designers should start by asking themselves what activity is under consideration and how they want the user to experience this activity, rather than jumping directly to the visual or interaction design.

The contradictory results we found across the studies focusing, e.g., on social supportiveness or visual and bodily presence of PA, suggest that the impact of PA is subject to complex moderation. We claim that there is no optimal generic design of a PA, but rather that the design aspects need to be chosen to offer an experience consistent with the activity and the desired outcome. Subfields of IS and computer science seem to engage in a race to cross the *uncanny valley*, i.e., to create a virtual character that evokes the same emotional response as a human would [169]. We ask if crossing the valley will generate responses that correlate with the object of an activity, e.g., learning, and improvement in this regard. Results from some studies attending to the visual or bodily presence do not support this thesis [2,24,141,149]. Instead, PA should adapt to the learner. Some studies on visual and bodily presence [95], but also those pertaining to learner's prior knowledge [95,130,179] or PA's gender [147,165] suggest it. Further, another category of studies attending to various adaptations of PA [53,122,156] suggests positive impacts of personalization on learning outcomes. This strengthens the intuition that learners perceive PA as a social being: they perform better if they are more likely to experience a PA that fits their preferences. This, however, does not necessarily imply that they need a PA that looks like a human and mimics human verbal performance. Instead, it might well be that some learners will exhibit better connection with a playful character, while others will benefit more with a simple chatbot or a very realistic virtual human. This has implications for the ongoing IS research on applications of DA: instead of chasing the expensive ideal of virtual human, the research should focus more on enabling a smooth and effortless adaptation to the user's preferences and to the situation at hand. This might involve easy approaches like giving the user a chance to choose a DA they want to use or hardwiring how the DA should behave to fit a specific task and range up to dynamic adaptivity based on assumptions about the state of the user or the conversation. Despite some recent progress in equipping DA with a so-called *theory of mind* to predict the intentions of the user [33,39,45], the research remains unattended by IS. Furthermore, a *theory of collaboration* for DA to interpret the state of a multi-party encounter is missing. We claim that improvements in both areas could potentially yield DAs that better align with the user and, ultimately, help them reach their objectives. Instead of aiming for increasingly human-like, anthropomorphic DA, IS should investigate the fit among the user, the DA, and the activity.

*Table 3. Summary of research and design challenges concerned with development and deployment of DA.*

| | |
|---|---|
| Production | - What is the difference between engineering a social encounter involving DA, designing a DA, and designing a conventional IT system?<br>- What design processes and practices can support designing a DA or a social encounter with a DA? How does this impact the skills and knowledge designers need to acquire during their education?<br>- What tools can empower people to design a DA or to engineer a social encounter involving a DA?<br>- What is the role of large NLP and AI platform providers in the process of developing a DA?<br>- How do designers perceive the creating a DA? How do existing technologies impact the way designers or users think about DA design? |
| Distribution | - What roles or tasks can be taken over by a DA? What roles or tasks should not and why?<br>- What activities can benefit from including a DA? What activities might be impaired by a DA?<br>- How will roles of humans change with the diffusion of DAs?<br>- What model or theory of collaboration is applicable to make a DA understand collaboration?<br>- Who, depending on the context, should act as the provider of the DA (e.g., teacher, school, no-one)? |
| Exchange | - What social norms and rituals are likely to change with the proliferation of DAs?<br>- How can one assess the quality of a DA or estimate its impact?<br>- What ethical norms should designers obey when designing a DA? When is a DA unethical?<br>- What model or theory of moral behavior is applicable to guide the decisions of a DA?<br>- What are the obstacles for the provision of DA to a broader public? What might impair the uptake of this technology?<br>- What business models can emerge around the provision of DAs to the broader public? |
| Consumption | - What are the longitudinal effects of using a DA? How can one study long-term effects of using a DA?<br>- How do the positive or negative effects studied in experiments change when DA is used longer?<br>- How can one make DA adaptable to various contexts to facilitate its usage?<br>- How should a DA transform during use? How should the acquaintance with the user(s) and the context change the behavior of the DA?<br>- What happens when a DA ends its duty for a user or group of users? Should the ontological status of an agent change when it is not used anymore |

Overall, the research on LPA points to a range of questions that go beyond the education context and need attention from the broader IS community. Table 3 provides an overview of challenges for research regarding development and deployment of DA. We systematize the issues along the four subsystems of an activity system: production, distribution, exchange, and consumption. We position the designer as subject, and the human-DA interaction as the object of their activity. This produces a research agenda which advances previous attempts to collect and structure directions for IS research [127,167,168]. Whereas some questions were asked before already, the AT framing yields questions regarding roles and skills necessary for the design of a DA (*production*), as well as the role of a "provider" included in the design of the activity (*distribution*). It also attends to the long-term effects of using a DA and how the usage, the user, and the community might change as time proceeds.

The questions as well as the results for PA indicate that we need a holistic perspective on DA. We hypothesize that the AT perspective on LPA can be applied to any activity involving a DA. Figure 8 provides an abstraction of the LPAM with potential for application beyond the context of education. We refer to it as the *activity with a digital agent model*. ADAM relies on the assumption that humans engage in complex or interdependent activities with DA based on a hypothetical image that DA is engaging in its activities in an independent way and is driven by some intentions. While this aligns with psychology studies suggesting that humans ascribe intentions and human attributes to artefacts [45,46,117,139] and even with recent research on perception of algorithms by users [159,176], the proposition goes beyond the idea of *objects with intent* proposed as an extension to AT elsewhere [159]. In our understanding, the DA is not object of the user's activity: it is not the agent that undergoes the transformation during the interaction (unlike in [159]). Instead, we claim that the user perceives DA as an agent whose desired outcomes pertain to but might not be identical with the objectives of the user in interdependent activity systems. The *ascribed agency assumption* extends the AT in two respects: First, it enables perceived activity systems, i.e., activity systems that do not necessarily describe how an activity emerges, but how an actor perceives the activity of another subject. By ascribing agency to this subject, the perceiver establishes a hypothetical image of the activity as it might be experienced by the subject. Second, it allows analyzing activity systems with a non-human and non-biological entity as a subject. However, it does not claim that activities described by those systems exist in the real world, but rather that they emerge within the mind of the perceiver. Nevertheless, understanding those hypothetical activities is important because they might drive real-world activities of the perceiver. AT is not a static theory but one which has been reinterpreted and updated over the last decades [41,72,115,120].

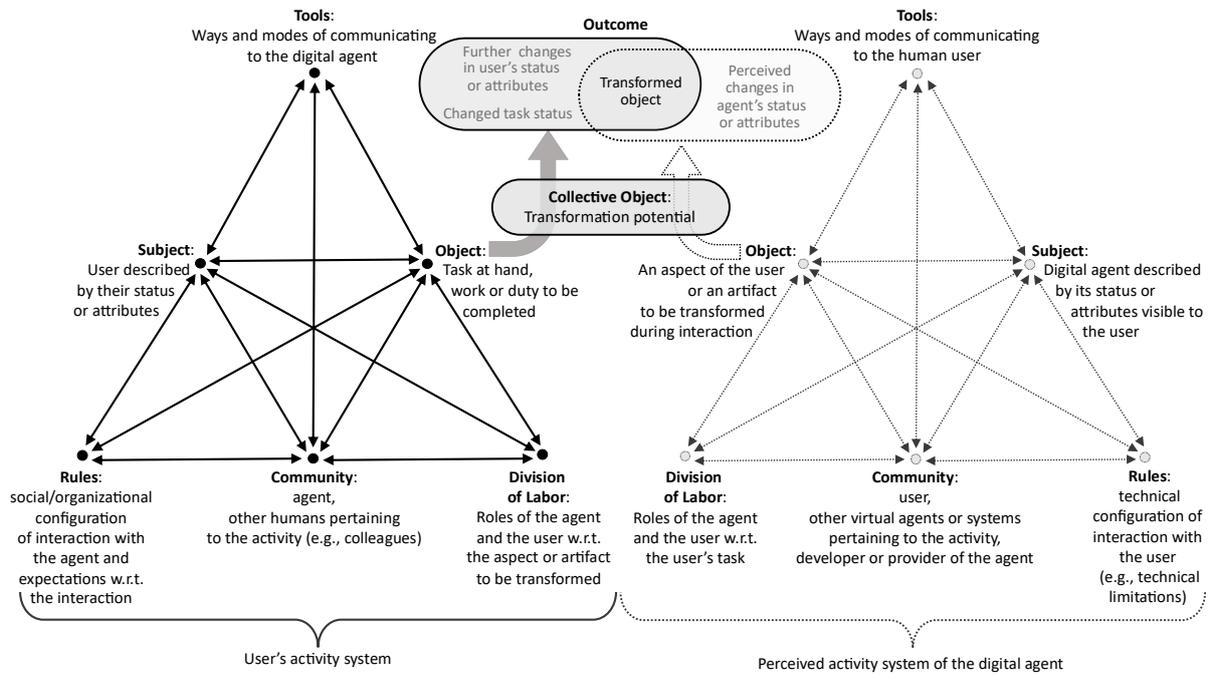

*Figure 4. ADAM ('activity with a digital agent' model): Model of two interdependent activity systems embracing interaction with a digital agent; generalization of LPAM beyond educational context.*

There have been calls for reconsidering AT with a focus on agency of non-biological beings [112,113,175,176]. This article responds to those calls by offering a solution that does not change the ontological status of artefacts – lacking agency or intentions on their own – but still allows analysis of complex activities or interrelated activity systems in which those artefacts invoke a sense of agency in an observer or members of the community. Application of this perspective is not only useful for classification of published studies or analysis of existing activities. We suggest that ADAM can be used in design to describe what a user should experience when engaging in an activity with a DA. On the one hand, it can help identify the broader context of the focal activity in its current form (community, rules, etc.). On the other hand, it points towards important aspects of DA design: what should the user think are the objects or desired outcomes of the DA, what should be the rules that the user perceives when interacting with the DA, how should the DA position itself towards other members of the community, and what division of labor should be suggested by the design? Earlier research documents that designing DA is a complex undertaking involving much creative work and attention to social dynamics [64]. In fact, we argue that designing DA is more about *engineering a social encounter* than designing a technology artefact. ADAM provides guidance for engineers approaching this complex task by pointing out the relevant aspects in a structured way. Overall, we argue that designing DA is more about engineering a social encounter than designing a technology artefact. Table 4 explicates the questions that emerge when a designer embraces ADAM as a framework for engineering a social encounter involving a DA. While it does not provide clear answers, it necessarily provides the right questions designers owe to ask when designing a DA.

*Table 4. Questions to guide the engineering of a social encounter involving a DA.*

| | |
|---|---|
| Production Subsystems (in user's activity system, AS, and in the perceived AS of DA) | - Who are the human actors interacting with the DA in the focal activity?<br>- What tasks do they conduct? What are their intermediate and ultimate objectives?<br>- In what ways should they interact with the DA? In what ways should the DA interact with them?<br>- As what or as whom should human actors perceive the DA?<br>- What objectives should the human actors ascribe to the DA?<br>- What aspects of the DA's design would help the humans identify DA's objectives? How does the DA communicate its objectives?<br>- How do the objectives of humans and the objectives ascribed to the DA interrelate?<br>- What should be the ultimate outcome of the collaboration between the DA and the human actors? |
| Distribution Subsystems (in user's AS and in the perceived AS of DA) | - What are the relationships between human actors within the focal activity?<br>- What are the social and organizational identities of humans involved in the activity?<br>- What are the relationships between the focal DA, other DAs, and further technology in the activity?<br>- What social or organizational identity should the human actors attribute to the DA?<br>- What aspects of the DA's design would help the humans identify DA's identity? How should the DA communicate its role?<br>- Whom should the user see as the provider / the person behind the DA? How should the DA communicate its relationship to human actors who might not participate directly in the activity?<br>- How are the tasks, roles, and responsibilities related to the object distributed between the human actors and the DAs? What aspects of the DA's design help humans understand their own role?<br>- How and to what extent can human actors and the DA negotiate their roles? |
| Exchange Subsystems (in user's AS and in the perceived AS of DA) | - What organizational or social norms govern the focal activity?<br>- What expectations do human actors have towards the focal activity?<br>- What expectations and norms should be instantiated through the design of the DA?<br>- What technical limitations govern the action of the DA?<br>- What aspects of the DA's design would help the humans identify DA's limitations? How does the DA communicate its limitations?<br>- What aspects of the DA's design would help the humans identify the desired norms or expectations? How does the DA communicate what norms it follows?<br>- How do technical limitations and organizational/social norms interrelate? |
| Consumption Subsystems (in user's AS and in the perceived AS of DA) | - What is the temporal and social order of the focal activity?<br>- How should the DA join the activity and detach from it?<br>- How should the status and the attributes of the DA and the human actors change through the transformation of the object? How do they evolve throughout the activity?<br>- How should the relationships among the human actors change throughout the activity?<br>- How should the relationship between the human actors and the DA change throughout the activity? |

**LIMITATIONS AND CONCLUSIONS**

The presented guidance not only summarizes the considered studies, but also extends, updates, and consolidates results, which were previously scattered across several literature reviews [9,50,88,113,141,148,200], and presents them in a an action-oriented fashion. However, the article has limitations related first to its empirical method, the systematic literature review. Because we consider only peer-reviewed, scientific publications, the analysis may suffer from publication bias, missing studies published in theses or dissertations. Also, the search methodology (e.g., considering titles, abstracts, and keywords only) and the selection process (e.g., strict exclusion criteria for the cooperative learning scenario) can introduce further biases. We hope we addressed those risks by conducting an extensive forward-and-backward search, and by designing the selection and classification process as a collaborative effort. Secondly, limitations result from the conceptual framework we developed. Choosing AT as a theoretical lens directs attention to the specific aspects of AT, such as the mediating nature of tools and communities in a learning activity. Also, the definition of learning outcome as the ultimate motive of a learning activity can be contested: many analyzed studies consider adjacent outcomes such as motivation or engagement, which suggests the learning activity can meet other distinct needs of the learner. Those limitations offer opportunities for further research oriented at (a) extending the LPAM to cooperative learning situations involving many learners, a PA, and human agents like a teacher, (b) adapting the LPAM to accommodate broader motives and outcomes beyond the learning outcome, and (c) testing the plausibility of the model with particular focus on the intentionality assumption that the learner perceives PA to be involved in activity of its own. Those aspects can be addressed in design studies and further meta-analyses or reviews.

Nevertheless, we see the current article as a step towards unpacking the role PA and, more generally, DA can take on. We see potential in adapting the LPAM to analyze commercial service agents, social chatbots, private virtual assistants, health assistants, and industrial applications of social robots. Further, IS researchers benefit from the identified research directions concerning pedagogical agents outlined in Section 5.1 – by addressing those gaps they can contribute to IS and to education studies. Pedagogy and education researchers receive a comprehensive analysis of past research related to PAs. They can benefit from a new, inclusive definition of a pedagogical agent that accommodates the technological improvements of the last decade and growth in the fields of social robotics and applied artificial intelligence. Identified links to core IS theories can be leveraged to position research beyond disciplinary borders. Finally, developers and designers, whether professionals or researchers, benefit from the discussion about various design aspects and an overview of past research provided in a simple and accessible form that may contribute to popularization of PAs. This will open doors for studying appropriation of those technologies in educational practice. Additionally, they get a model of activities involving interaction with a DA, ADAM, to guide their design and analysis efforts.

## FIGURES AND TABLES (IN ORDER OF APPEARANCE)

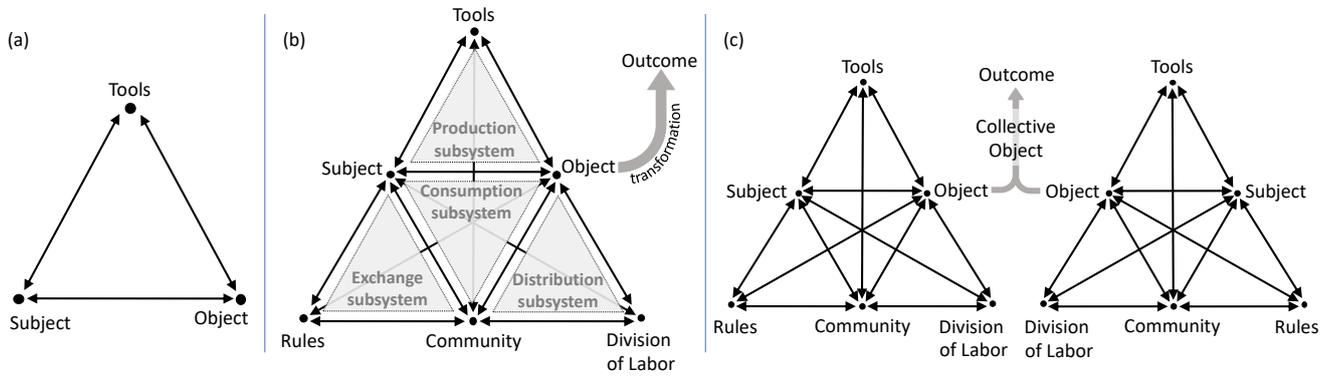

*Figure 5. Three generations of activity theory: (a) an activity system according to the first generation AT as devised by Vygotsky [183], (b) second generation AT as proposed by Leontyev [120] and represented by Engeström [68] including the subsystems of an activity system, (c) third generation AT as proposed by Engeström [68,70,72] for describing interdependent activity systems.*

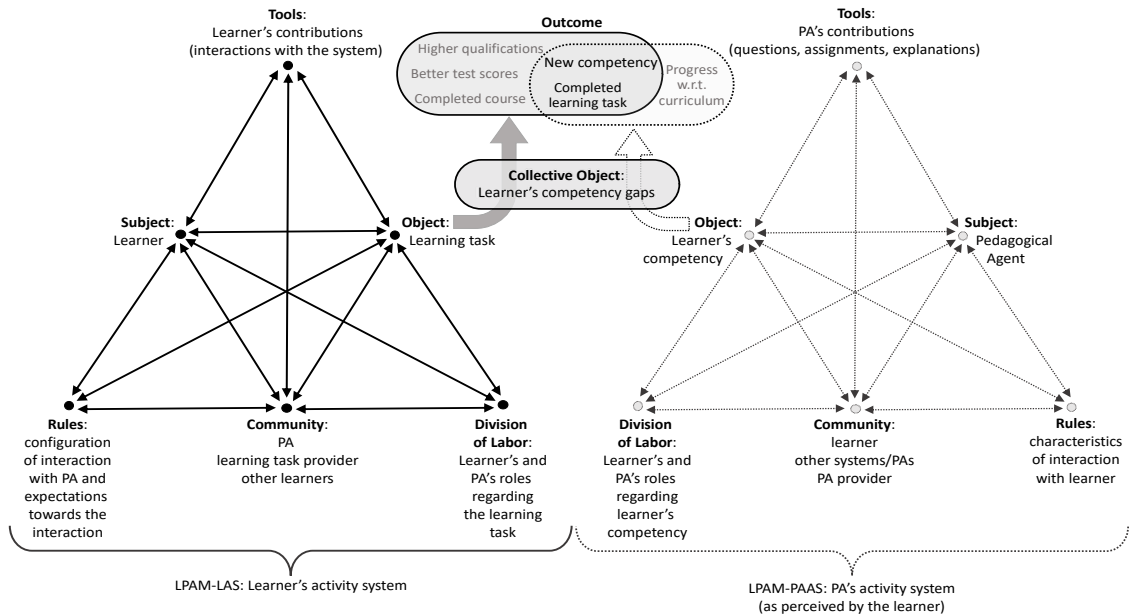

*Figure 6. LPAM: Model of interdependent activity systems involved in learning with a PA based on Engeström's conception of interdependent activity systems [68,70,72] and Dennett's theory of intentionality [46]*

*Table 5. Number of articles by learner subject category*

| Learner Subject Category | Primary school | Secondary school | Higher education | Adults |
|---|---|---|---|---|
| **Number of projects** | 7 | 7 | 46 | 4 |

*Table 6. Number of articles by PA subject category. If more than one category was tested, we count the paper as belonging to the more advanced category*

| PA Subject Category | Chatbot | Cartoon agent (e.g., avatar) | Human-like agent (e.g., virtual human) | Social Robot |
|---|---|---|---|---|
| **Number of projects** | 10 | 11 | 35 | 8 |

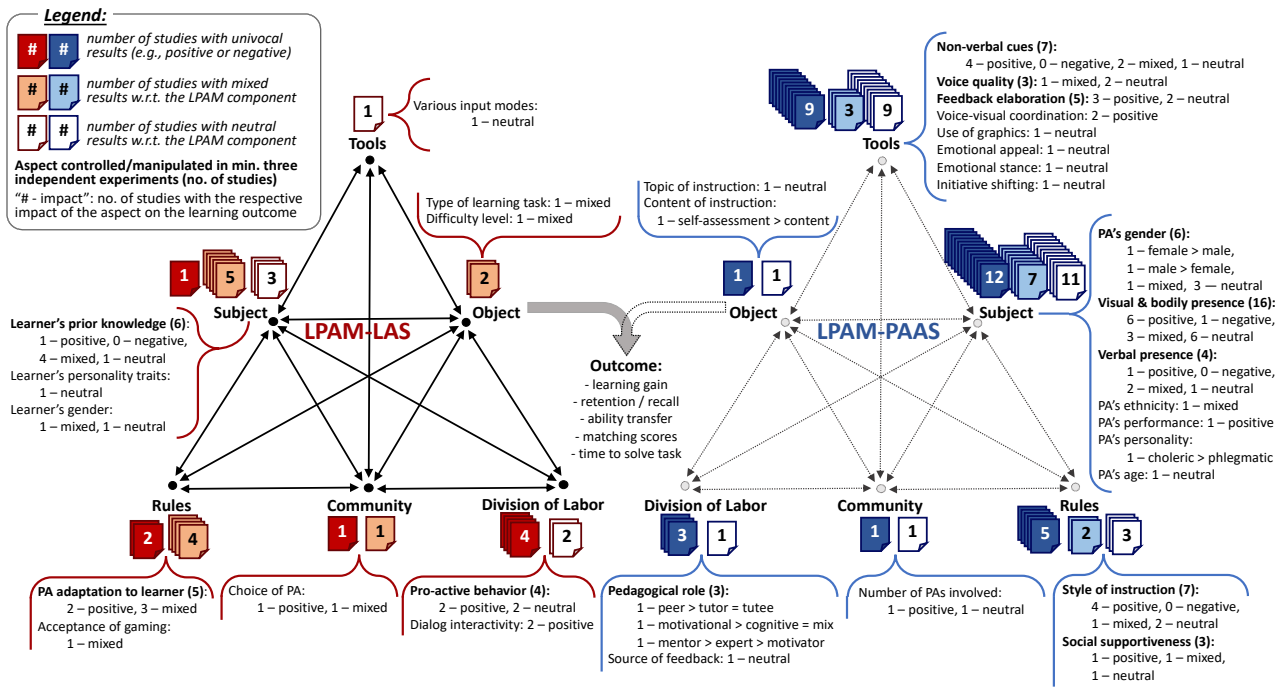

*Figure 7. Results overview of the literature review depicted in the LPAM including relevant design aspects.*

*Table 7. Summary of research and design challenges concerned with development and deployment of DA.*

| | |
|---|---|
| Production | - What is the difference between engineering a social encounter involving DA, designing a DA, and designing a conventional IT system?<br>- What design processes and practices can support designing a DA or a social encounter with a DA? How does this impact the skills and knowledge designers need to acquire during their education?<br>- What tools can empower people to design a DA or to engineer a social encounter involving a DA?<br>- What is the role of large NLP and AI platform providers in the process of developing a DA?<br>- How do designers perceive the creating a DA? How do existing technologies impact the way designers or users think about DA design? |
| Distribution | - What roles or tasks can be taken over by a DA? What roles or tasks should not and why?<br>- What activities can benefit from including a DA? What activities might be impaired by a DA?<br>- How will roles of humans change with the diffusion of DAs?<br>- What model or theory of collaboration is applicable to make a DA understand collaboration?<br>- Who, depending on the context, should act as the provider of the DA (e.g., teacher, school, no-one)? |
| Exchange | - What social norms and rituals are likely to change with the proliferation of DAs?<br>- How can one assess the quality of a DA or estimate its impact?<br>- What ethical norms should designers obey when designing a DA? When is a DA unethical?<br>- What model or theory of moral behavior is applicable to guide the decisions of a DA?<br>- What are the obstacles for the provision of DA to a broader public? What might impair the uptake of this technology?<br>- What business models can emerge around the provision of DAs to the broader public? |
| Consumption | - What are the longitudinal effects of using a DA? How can one study long-term effects of using a DA?<br>- How do the positive or negative effects studied in experiments change when DA is used longer?<br>- How can one make DA adaptable to various contexts to facilitate its usage?<br>- How should a DA transform during use? How should the acquaintance with the user(s) and the context change the behavior of the DA?<br>- What happens when a DA ends its duty for a user or group of users? Should the ontological status of an agent change when it is not used anymore |

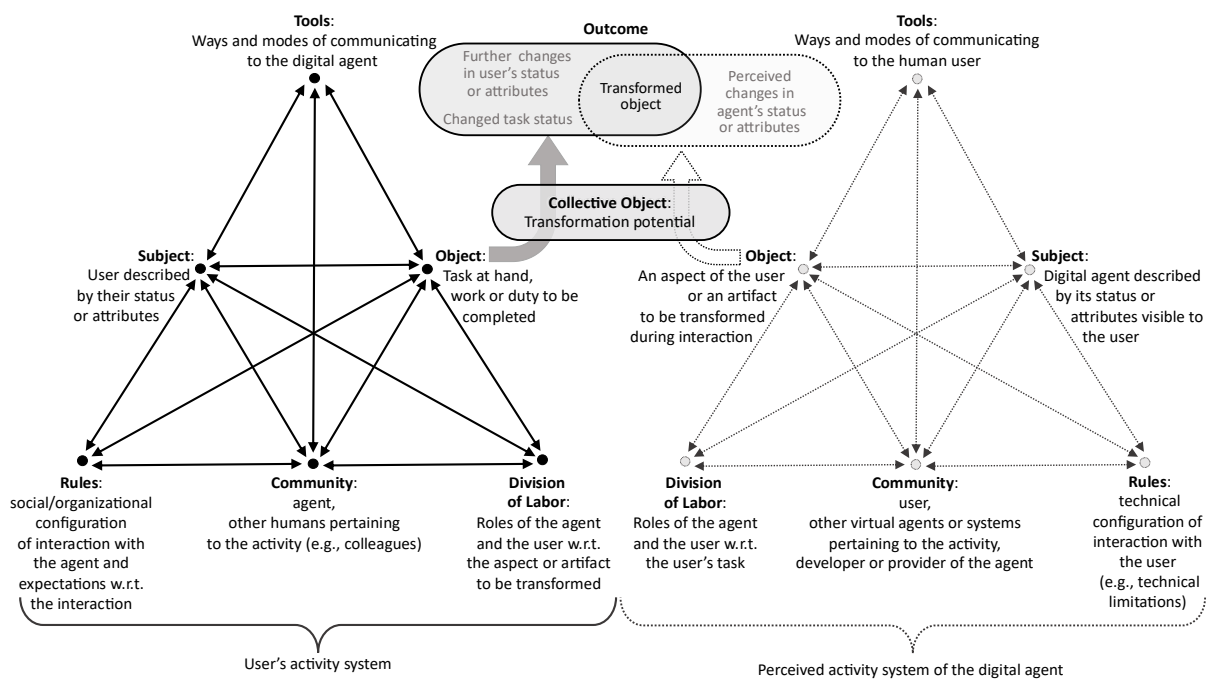

*Figure 8. ADAM ('activity with a digital agent' model): Model of two interdependent activity systems embracing interaction with a digital agent; generalization of LPAM beyond educational context.*

*Table 8. Questions to guide the engineering of a social encounter involving a DA.*

| | |
|---|---|
| *Production Subsystems (in user's activity system, AS, and in the perceived AS of DA)* | - Who are the human actors interacting with the DA in the focal activity?<br>- What tasks do they conduct? What are their intermediate and ultimate objectives?<br>- In what ways should they interact with the DA? In what ways should the DA interact with them?<br>- As what or as whom should human actors perceive the DA?<br>- What objectives should the human actors ascribe to the DA?<br>- What aspects of the DA's design would help the humans identify DA's objectives? How does the DA communicate its objectives?<br>- How do the objectives of humans and the objectives ascribed to the DA interrelate?<br>- What should be the ultimate outcome of the collaboration between the DA and the human actors? |
| *Distribution Subsystems (in user's AS and in the perceived AS of DA)* | - What are the relationships between human actors within the focal activity?<br>- What are the social and organizational identities of humans involved in the activity?<br>- What are the relationships between the focal DA, other DAs, and further technology in the activity?<br>- What social or organizational identity should the human actors attribute to the DA?<br>- What aspects of the DA's design would help the humans identify DA's identity? How should the DA communicate its role?<br>- Whom should the user see as the provider / the person behind the DA? How should the DA communicate its relationship to human actors who might not participate directly in the activity?<br>- How are the tasks, roles, and responsibilities related to the object distributed between the human actors and the DAs? What aspects of the DA's design help humans understand their own role?<br>- How and to what extent can human actors and the DA negotiate their roles? |
| *Exchange Subsystems (in user's AS and in the perceived AS of DA)* | - What organizational or social norms govern the focal activity?<br>- What expectations do human actors have towards the focal activity?<br>- What expectations and norms should be instantiated through the design of the DA?<br>- What technical limitations govern the action of the DA?<br>- What aspects of the DA's design would help the humans identify DA's limitations? How does the DA communicate its limitations?<br>- What aspects of the DA's design would help the humans identify the desired norms or expectations? How does the DA communicate what norms it follows?<br>- How do technical limitations and organizational/social norms interrelate? |
| *Consumption Subsystems (in user's AS and in the perceived AS of DA)* | - What is the temporal and social order of the focal activity?<br>- How should the DA join the activity and detach from it?<br>- How should the status and the attributes of the DA and the human actors change through the transformation of the object? How do they evolve throughout the activity?<br>- How should the relationships among the human actors change throughout the activity?<br>- How should the relationship between the human actors and the DA change throughout the activity? |

# Virtual Appendix A1

This virtual appendix belongs to the publication 'Learning with Digital Agents: An Analysis based on the Activity Theory' by Dolata et al. published in the Journal of Management Information Systems in 2023.

Table A1 provides a summary of each study considered in the literature review that forms the basis for the study. Whereas some variables or aspects appear in a single study only (e.g., learner's gender, see the first row), others are considered by multiple articles (e.g., learner's prior knowledge, see the rows 4-10).

All variables from Table A1 are placed in relation to one of the *learning with a pedagogical agent model* (LPAM) components depicted in Figure 2 provided in the article. To clarify what aspect belongs to which activity theory (AT) system and component, we used color-coding in the first two columns. Dark red in the first column indicates components belonging to *learning with a pedagogical agent model – learner's activity system* (LPAM-LAS), dark blue – components belonging to *learning with a pedagogical agent model – pedagogical agent's activity system* (LPAM-PAAS). Each component is marked with a pale color in the second column: subject with pale violet, object with pale blue, tools with pale green, etc. If at least three independent studies attend to the same design aspect, we listed them one after another and marked the whole group grey. This shows which design aspects were studied most frequently.

*Table A1. Summary of results regarding the impact of LPAM's components on learning outcome. The highlighted fields form groups of at least three studies addressing a similar design aspect.*

| LPAM Component | | Manipulated or controlled variable (learning task) [Reference] | Impact of the manipulated/controlled variable on the learning outcomes |
|---|---|---|---|
| LPAM-LAS | LAS: Subject | **Learner's gender**: female vs. male (learning physics) [van der Meij et al. 2015] | **neutral**; no effect of learner's gender on the learning gain as measured by pre- and post-test |
| | | **Learner's gender**: female vs. male (listening to a story to be recalled later) [Mutlu et al. 2006] | **mixed**; only females perform significantly better if the PA looks at them 80% of time as compared to when PA looks at them only 20% of time |
| | | **Learner's personality traits**: agreeableness vs. conscientiousness vs. neuroticism vs. extraversion vs. openness to experience as assessed in a survey (learning about human circulatory system) [Harley et al. 2016] | **neutral**; no correlation between learner's personality traits and learning outcome in learning with a PA as measured by pre- and post-test; complex interactions with emotional response towards PA |
| | | **Learner's prior knowledge**: low vs. high level (learning electrical circuits analysis) [Johnson et al. 2013] | **mixed**; low-level learners have higher learning gains if learning with a PA using signaling cues (pointing), no such effect for high-level learners |
| | | **Learner's prior knowledge**: low vs. high level (mastering chemical stoichiometry) [McLaren et al. 2011] | **neutral**; all learners benefit from politely or directly communicating PA despite their level |
| | | **Learner's prior knowledge**: low vs. middle vs. high (mastering programming problem solving) [Hooshyar et al. 2015] | **mixed**; low- and middle-level learners make larger learning gains than the high-level ones do |
| | | **Learner's prior knowledge**: low vs. middle vs. high (learning foreign-language grammar) [Choi and Clark 2006] | **positive**; low-level learners have significantly higher learning gains than middle-level ones, and middle-level ones higher than the high-level ones when using a PA |
| | | **Learner's prior knowledge**: above-average vs. below-average learner (memorizing multiplication table) [Konijn and Hoorn 2020] | **mixed**; above-average learners have higher learning gains with a socially supportive PA while below-average learners reach higher gains with a neutral PA as measured by pre- and post-test |
| | | **Learner's prior knowledge**: novice vs. intermediate learner (learning qualitative physics) [VanLehn et al. 2007] | **mixed**; no significant difference if the learner's level matches the level of the content to be learned; if the content is above learner's level, PA-based tutoring outperforms learning without a PA |
| | LAS: Object | **Type of learning task**: procedural learning – learning how to perform a task with a software vs. attitudinal learning – learning to make a decision in alignment with specific values (improving MS Office skills and deciding about intellectual property) [Baylor and Kim 2009] | **mixed**; no direct effect on the learning outcome as measured by a post-test; effects moderated by facial expression and deictic gesture of the PA –learning gains are higher in attitudinal learning when PA's facial expression is present and deictic gesture is absent, and in procedural learning when PA's deictic gesture is present |
| | | **Difficulty level of the content**: intermediate vs. novice content (learning qualitative physics) [VanLehn et al. 2007] | **mixed**; no significant difference if the learner's level matches the level of the content to be learned; if the content is above learner's level, PA-based tutoring outperforms learning without a PA |
| | LAS: Tools | **Various input modes**: speech-based input in dialog with PA vs. text-based input in dialog with PA vs. no interaction with PA (learning basics of computer science) [D'Mello et al. 2010] | **neutral**; no significant difference between speech-based and text-based input mode conditions w.r.t. learning gain as by pre- and post-test, but broader topic coverage in speech-based input condition |
| | LAS: Com- | **Choice of PA**: Learners having choice between different archetypes of PA, e.g., between young male PA and old female PA, vs. learners without choice | **positive**; learners with choice options had better learning gains in a delayed post-test (but not immediate post-test) |
| | | (learning electrical circuit analysis) [Ozogul et al. 2013] | than learners without option to choose from the available PA archetypes |
| | | **Choice of PA**: Learners having choice between different PAs in terms of gender and ethnicity vs. learners without choice (learning about car mechanics) [Moreno and Flowerday 2006] | **mixed**; no significant impact of the condition on the learning gains of learners as measured by pre- and post-test; learners who can choose and choose a PA with different ethnicity than their own outperform other learners |
| | LAS: Rules | **PA adaptation** to fit the learner's learning style: adapting PA vs. neutral-adapting PA vs. mismatching PA (training SQL skills) [Latham et al. 2014] | **mixed**; the learners in the mismatch condition have significantly worse results as measured by answers provided during the interaction with PA, but no differences concerning learning gains |
| | | **PA adaptation** to fit the learner's pause-related needs: reward strategy – breaks timing adapted to performance gains, vs. refocus strategy – breaks timing adapted to performance losses, vs. fixed timing (solving mathematical problems) [Ramachandran et al. 2017] | **positive**; learning gains in learners with a personalized break strategy, i.e., in reward and refocus strategy conditions, were significantly higher than the learning gains in the fixed timing condition |
| | | **PA adaptation** to the learner's disengagement level: PA reacting to learner's disengagement based on eye tracking vs. PA without adaptation capabilities (learning biology) [D'Mello et al. 2012] | **mixed**; learners in adaptive-PA condition had higher learning gains in deep-reasoning test, but lower learning gains in fact-based assertion test compared to non-adaptive-PA-condition learners |
| | | **PA adaptation** to the recognized learner's uncertainty: PA always reacting to the learner's uncertainty vs. PA not reacting at all vs. PA reacting in random manner (learning physics) [Forbes-Riley and Litman 2011] | **mixed**; learners in reactive-PA condition outperform the random-condition learners in terms of learning gains, but not the non-reactive PA |
| | | **PA adaptation** to the learner's skills: PA personalized by means of a Bayesian algorithm vs. PA personalized by an additive algorithm vs. non-personalized PA using random content for training vs. no training (solving the nonogram puzzle) [Leyzberg et al. 2014] | **positive**; learners in the adaptive conditions outperform other learners in terms of time needed for solving the puzzle in the second game; learners in the Bayesian-algorithm condition outperform the additive-algorithm condition in the third game |
| | | **Acceptance of gaming behaviors**: dialog PA accepting nonsense answers, proceeding after many nonsense answers – accepting a "gaming" behavior vs. dialog PA accepting only answers fitting a template vs. PA which "reads" information to the learner (learning research methods in psychology) [Hastings et al. 2010] | **mixed**; if students can choose the condition, the dialog PA accepting nonsense answers outperforms the reading PA in terms of learning gains; if students cannot choose the condition, the reading PA outperforms dialog-oriented PA which does not allow for nonsense answers |
| | LAS: Division of Labor | **PA's pro-active behavior**: Collaborative, prompt-and-feedback-based goal specification vs. pre-specified learning goals (learning about human circulatory system) [Harley et al. 2018] | **positive**; learners who could specify their learning goals collaboratively with the PA obtained higher learning gains compared to learners whose learning goals were pre-specified by the PA |
| | | **PA's pro-active behavior**: Collaborative, prompt-and-feedback-based goal specification vs. pre-specified learning goals (learning about human circulatory system) [Harley et al. 2016] | **neutral**; no difference between conditions with regard to the learning gains as measured by a pre- and post-test |
| | | **PA's pro-active behavior**: Pro-active prompt-and-feedback guidance vs. on-demand guidance when requested by the learner (learning about human circulatory system) [Duffy and Azevedo 2015] | **neutral**; learners who received prompt-and-feedback guidance performed as well as learners in the on-demand guidance condition in terms of comprehension and achievement scores |
| | | **PA's pro-active behavior**: learners receive a pre-question to guide their learning episode vs. learners receive no pre-question<br>(learning about electric motors)<br>[Mayer et al. 2003] | **positive**; learners in the pre-question condition achieved better learning outcomes as measured by a problem-solving post-test compared to the condition without a pre-question |
| | | **Dialog interactivity**: interactive – learners can ask questions and receive answer vs. non-interactive – information presented as a multimedia message<br>(learning about electric motors)<br>[Mayer et al. 2003] | **positive**; learners in the interactive condition achieved better learning outcomes as measured by a problem-solving post-test compared to the non-interactive condition |
| | | **Dialog interactivity**: full – learner can provide multiple answers and receives feedback/hints to each answer, vs. limited – only one answer receives feedback, vs. mixed – full dialog in the first half-time, then limited, vs. no-dialog – no answers can be given<br>(training scientific inquiry skills)<br>[Kopp et al. 2012] | **positive**; the learners in the full and mixed-dialog conditions outperform the ones in no-dialog conditions; full and mixed-dialog exhibit similar learning gains as measured by pre- and post-test, but mixed-dialog condition is more efficient because it takes significantly less time to complete |
| **LPAM-PAAS** | **PAAS: Subject** | **Agent ethnicity as compared to learner's ethnicity**: similar vs. not similar<br>(learning about car mechanics)<br>[Moreno and Flowerday 2006] | **mixed**; no direct impact of the ethnical similarity on the learning gains; if the learners can choose PA, learners who choose an agent of different ethnicity perform better |
| | | **Agent gender**: same vs. opposite of the learner's gender<br>(solving mathematical tasks)<br>[Krämer et al. 2016] | **mixed**; if the PA of opposite gender displays rapport (nods, smiles), it positively impacts the learning outcome of the learner, otherwise no impact |
| | | **Agent gender**: same vs. opposite compared to the learner's gender<br>(learning about car mechanics)<br>[Moreno and Flowerday 2006] | **neutral**; no significant impact of the gender similarity on the learning gain as measured by a multiple-choice pre- and post-test |
| | | **Agent gender**: female vs. male<br>(learning electrical circuits analysis)<br>[Ozogul et al. 2013] | **female > male**; female and male learners both had higher learning gains when using a female PA |
| | | **Agent gender**: female vs. male<br>(learning about multimedia learning theory)<br>[Schroeder and Adesope 2015] | **neutral**; no difference between the conditions as measured by pre- and post-test |
| | | **Agent gender**: female vs. male<br>(designing an e-learning class for children)<br>[Kim et al. 2007] | **neutral**; no difference between the conditions as measured by an information recall post-test |
| | | **Agent gender**: female vs. male<br>(learning methods to design an e-learning class)<br>[Kim et al. 2007] | **male > female**; all learners had better outcomes in an information-recall test after using a male PA |
| | | **The problem-solving performance of the PA as a co-learner**: high vs. low performing agent<br>(mastering the morse code)<br>[Ju et al. 2005] | **positive**; learners working with high-performing co-learner PA made significantly fewer errors in a comprehension test than other learners |
| | | **Visual and bodily presence**: physical presence as a non-humanoid robot vs. video representation of the PA on a screen vs. voice-only PA vs. no PA<br>(solving nonogram puzzle)<br>[Leyzberg et al. 2012] | **positive**; learners in the physical-presence condition outperform all other learners in terms of time needed for solving the puzzle; all PA conditions outperform the no-PA condition |
| | | **Visual and bodily presence**: physical presence as a humanoid robot vs. voice-based narration without physical presence of a robot<br>(learning about prime numbers)<br>[Kennedy et al. 2015] | **positive**; the learning gain, i.e., difference between pre- and post-test, in condition with physically present robot is significant, while in the voice-based condition the gain is not significant |
| | | **Visual and bodily presence**: text only vs. voice only vs. visible avatar that delivers instruction by voice with no gestures vs. visible avatar that delivers | **mixed**; all conditions except for the text-only condition exhibit significant and comparably large learning gain as expressed by pre- and post-test, but all conditions in |

| | |
|---|---|
| instruction by voice as well as movements, gestures, and gazes<br>(improving foreign-language grammar)<br>[Carlotto and Jaques 2016] | which the PA communicates via voice reach similar effect |
| **Visual and bodily presence**: voice-based narration and a visible avatar vs. voice-based narration vs. text only<br>(learning about cardiovascular system)<br>[Dunsworth and Atkinson 2007] | **positive**; learners in the condition involving voice-based narration and a visible avatar outperform learners in the two other conditions as measured by a post-test; no difference between the voice-based and text-only narration |
| **Visual and bodily presence**: animated and dynamic avatar and voice-based narration featuring coordinated motion, gesture, gaze, and voice, vs. static avatar and voice-based narration vs. voice-only narration<br>(solving mathematical problems)<br>[Lusk and Atkinson 2007] | **positive**; learners in the animated PA condition outperformed learners in the voice-only condition in terms of near and far transfer post-test performance; no difference between the static PA condition and the other two |
| **Visual and bodily presence**: human-voice narration and avatar image vs. personalized text on the screen and no avatar image vs. no narration<br>(learning about intellectual property)<br>[Park 2015] | **neutral**; no impact on the learning outcome as measured by the recall and comprehension post-test |
| **Visual and bodily presence**: text only vs. speech only vs. talking head, i.e., animated avatar and voice-output, vs. talking head and text<br>(learning computer science basics)<br>[Graesser et al. 2003] | **neutral**: no direct effect of the medium on the learning performance as measured by the knowledge pre- and post-test |
| **Visual and bodily presence**: PA represented by an animated avatar vs. static avatar vs. no avatar<br>(preparing an instructional plan)<br>[Baylor and Ryu 2003] | **neutral**; no difference between conditions according to a post-hoc assessment of the instructional plans prepared during the interaction with the PA |
| **Visual and bodily presence**: PA represented by image and voice vs. text-only explanation and guidance<br>(solving word problems)<br>[Atkinson 2002] | **positive**; learners in the voice-and-visual condition perform better than the learners in the text-only condition in an ability transfer post-test |
| **Visual and bodily presence**: PA represented by image and voice vs. only voice vs. no image and no voice<br>(learning physics)<br>[van der Meij et al. 2015] | **neutral**; no impact of the condition on the learning outcome as measured by pre- and post-test |
| **Visual and bodily presence**: Two visualized PAs (tutor and co-learner) in a chat trialogue with the learner vs. text in pop-up callouts without a visualized PA<br>(practicing algebra calculations)<br>[Nye et al. 2018] | **mixed**; neutral for learning gain as measured by a pre- and non-delayed post-test; positive for a delayed basic skill diagnostic test several weeks after the treatment – learners from the trialogue condition outperform others |
| **Visual and bodily presence**: Animated avatar with coordinated verbal and non-verbal behaviors vs. an animated hand with a small subset of non-verbal behavior vs. human teacher<br>(learning foreign language idioms)<br>[Ahmadi et al. 2017] | **negative**; learners interacting with the animated hand performed best, followed by those interacting with anthropomorphic PA, and the ones in teacher treatment performed worst as measured by the pre- and post-test measurement of learning gain |
| **Visual and bodily presence**: Animated avatar vs. an animated pointing hand vs. no visual representation<br>(understanding neurotransmission)<br>[Wang et al. 2018] | **positive**; learners in the animated-avatar condition outperform learners in the other two conditions as measured by retention and transfer test |
| **Visual and bodily presence**: avatar using gesture to signal relevance vs. arrow signaling vs. no signaling<br>(learning electric circuits analysis)<br>[Johnson et al. 2013] | **mixed**; learners with low-level knowledge in avatar condition outperform other low-level learners, no effect on high-level learners |

| | | **Visual and bodily presence**: Animated avatar of the PA vs. a simple arrow; both with a voice overlay (learning foreign-language grammar) [Choi and Clark 2006] | **neutral**; no influence of the condition on the learner's gains as measured by pre- and post-test |
|---|---|---|---|
| | | **Visual and bodily presence**: PA represented as a virtual human with synchronized voice vs. voice only (understanding cognitive load theory) [Schroeder 2017] | **neutral**; no significant effect on the learning gain as measured by pre- and post-testing |
| | | **Personality of PA**: phlegmatic vs. choleric (learning tai chi martial arts) [Bian et al. 2016] | **choleric > phlegmatic**; learners in choleric condition show higher learning performance than others |
| | | **Verbal presence**: voice narration and text vs. text-only (mastering chemical stoichiometry) [McLaren et al. 2011] | **neutral**; no impact of the medium on the learning gains as measured by pre- and post-testing |
| | | **Verbal presence**: voice narration vs. text-only (learning about electric motors) [Mayer et al. 2003] | **positive**; learners in voice condition outperform others in a problem-solving transfer post-test |
| | | **Verbal presence**: voice narration with text vs. voice-narration-only vs. text-only (learning about lightning) [Craig et al. 2002] | **mixed**; learners in the voice-narration-only outperformed learners in the two other conditions in a retention, matching, and transfer pre- and post-test |
| | | **Verbal presence**: voice+text vs. text-only vs. no-PA (learning python programming) [Winkler et al. 2020] | **mixed**; voice-+text learners outperform no-PA learners in a transferability test, no difference to text-only |
| | | **PA's age**: PA represented as old vs. young by means of visual representation and a priming text (learning facts about allergies) [Beege et al. 2017] | **neutral**; no general effect of the perceived age on information retention or knowledge transfer |
| | **PAAS: Object** | **Topic of instruction**: nutrition, exercising, lifestyle (controlling own weight) [Brust-Renck et al. 2017] | **neutral**; the impact of the PA on the learning outcome was the same across topics |
| | | **Content of instructions**: directly referring to learner's ability to self-assess her competency vs. instructions oriented at the content (solving first-level equations) [Kautzmann and Jaques 2019] | **positive**; learners who learned with PA referring to learner's metacognition abilities exhibit higher performance in solving equations as measured by a pre- and post-test |
| | **PAAS: Tools** | **Non-verbal cues**: Gesturing vs. no gesturing (solving first-level equations) [Cook et al. 2017] | **positive**; learners who used gesturing avatar solve more tasks correctly and more quickly |
| | | **Non-verbal cues**: Gesturing vs. no gesturing (understanding information on formation of lightning in a foreign language) [Davis and Vincent 2019] | **positive**; learners who learned with gesturing avatar are more effective at recalling the content of the instruction in an information retention test |
| | | **Non-verbal cues**: Gesturing vs. no gesturing (understanding neurotransmission) [Wang et al. 2018] | **positive**; learners presented a gesturing avatar outperformed others as measured by retention and transfer test |
| | | **Non-verbal cues**: Deictic gesturing vs. no gesturing (assessment of MS Office skills and deciding about intellectual property) [Baylor and Kim 2009] | **mixed**; moderated by the type of learning task – deictic-gesture condition outperformed no-gesturing in procedural but not in attitudinal learning |
| | | **Non-verbal cues**: avatar visible using deictic gesturing vs. avatar visible and no gesturing vs. no avatar (learning about lightning) [Craig et al. 2002] | **neutral**; gestures or visualization of the agent did not impact the learning gains as measured in a retention, matching, and transfer pre- and post-test |
| | | **Non-verbal cues**: PA with dynamic facial expression vs. PA with neutral, static facial expression | **mixed**; impact on the learning outcome as measured by a post-test was moderated the learning task type – dynamic-facial-expression condition outperformed the |

| | | |
|---|---|---|
| | (assessment of MS Office skills and deciding about intellectual property) [Baylor and Kim 2009] | other condition in attitudinal but not in procedural learning |
| | **Non-verbal cues**: PA looking at the learner 80% of the time vs. 20% of the time when telling a story (listening to a story to be recalled later) [Mutlu et al. 2006] | **positive**; learners in the 80% condition performed significantly better during an open-ended recall post-test than the 20%-condition learners |
| | **Coordination between visual and voice output**: dynamic – content presentation coordinated with narration vs. static – content is presented at once (solving mathematical problems) [Lusk and Atkinson 2007] | **positive**; learners in the dynamic condition outperform learners in the static condition as measured by the near and far ability transfer post-test |
| | **Coordination between visual and voice output**: animated images coordinated with the narration vs. highlighting coordinated with the narration vs. static (learning about lightning) [Craig et al. 2002] | **positive**; learners in animated and highlighted condition outperform learners in the static-images condition in a retention and matching pre- and post-test |
| | **Use of graphics**: images in addition to a textual description of a content vs. no images (learning about intellectual property) [Park 2015] | **neutral**; presenting an image in addition to the explanatory text does not impact the learning outcome in recall and comprehension tests |
| | **Voice quality**: low-quality text-to-speech engine vs. high-quality text-to-speech engine with advanced modulation and prosody vs. human voice (learning about formation of lightning) [Chiou et al. 2020] | **neutral**; no significant difference between treatments in a post-test using multiple-choice questionnaire, an information retention test, and an ability transfer test |
| | **Voice quality**: classic text-to-speech vs. modern text-to-speech vs. human voice (learning about formation of lightning) [Craig and Schroeder 2017] | **mixed**: learners using PA with a modern text-to-speech outperformed other learners measured by recall, retention, and transfer tests |
| | **Voice quality**: weak-prosodic human voice vs. strong-prosodic human voice vs. text-to-speech engine (learning informatics in foreign language) [Davis et al. 2019] | **neutral**; no significant difference between treatments in an information-retention test |
| | **Emotional appeal**: Humor appeal vs. fear appeal vs. humor and fear appeal vs. no appeal (aviation safety training) [Buttussi and Chittaro 2020] | **neutral**; the appeals did not have an impact on the effectiveness of learning |
| | **Feedback elaboration**: elaborate feedback – providing and explaining the assessment vs. simple feedback – marking answer as correct or not (learning physics) [Lin et al. 2013] | **positive**; the learners in the elaborate-feedback condition outperform learners in the simple-feedback condition in terms of a post-test assessment of the learning outcome |
| | **Feedback elaboration**: full feedback after each answer with positive and procedural feedback vs. short hints only when learner's answer is wrong (translation of sentences into first order logic) [Perikos et al. 2017] | **positive**; learners who have learned with a PA providing full feedback show more improvement in the first-order-logic translation task than learners offered a PA with bottom-out hints |
| | **Feedback elaboration**: knowledge-construction dialog vs. question-answer dialog vs. dialog with a human tutor vs. no dialog (learning qualitative physics) [VanLehn et al. 2007] | **neutral**; direct comparison between the conditions does not exhibit any significant difference; difference moderated by the prior knowledge of the learner and the level of the content to be learned |
| | **Feedback elaboration**: scaffolding dialog vs. non-scaffolding dialog (learning python programming) [Winkler et al. 2020] | **positive**; learners in the scaffolding condition outperform learners in non-scaffolding condition in terms of information retention |

| | | | |
|---|---|---|---|
| | | **Feedback elaboration**: feedback oriented at self-regulatory learning, vs. normal task guidance only (learning about human circulatory system) [Harley et al. 2016] | **neutral**; no direct significant impact from such guidance and feedback on the learning outcome |
| | | **Emotional stance**: positive vs. negative vs. neutral (learning how to design e-learning class) [Kim et al. 2007] | **neutral**; no significant impact of the PA's emotional expression on the learners' outcome in a recall test |
| | | **PA's initiative shifting behavior**: utterances motivating learner to take over the initiative vs. neutral (learning programming skills) [Howard et al. 2017] | **neutral**; no significant impact of the PA's initiative shifting behavior on learning gain as measured by a knowledge pre- and post-test |
| | PAAS: Community | **Number of PAs involved in training**: Two animated PAs representing blood cells plus one anthropomorphic avatar vs. two animated blood cells only (learning about cardiovascular system) [Yung and Paas 2015] | **positive**; learners who were trained in the treatment including the anthropomorphic, pointing agent demonstrate better learning outcomes as measured by a comprehension test |
| | | **Number of PAs involved in training**: One vs. multiple (programming using MS Excel) [Dinçer and Doğanay 2017] | **neutral**; no significant difference between conditions involving a single or multiple PAs |
| | PAAS: Rules | **Style of instruction**: informal – personal, vs. formal (preparation for biology test) [Lin et al. 2020] | **positive**; informal style (e.g., directly addressing the learner) lead to better knowledge retention |
| | | **Style of instruction**: informal – personal, vs. formal (learning about ecology) [Brom et al. 2017] | **neutral**; no impact of the style on the learning outcomes in a retention and ability transfer post-test |
| | | **Style of instruction**: enthusiastic – smiling, nodding, positive, vs. neutral – calm, no nodding, no smiling (programming using C) [Liew et al. 2017] | **positive**; learners interacting with the enthusiastic PA perform better in a comprehension post-test |
| | | **Style of instruction**: enhanced rapport behavior, i.e., smiling, nodding, vs. no-rapport behavior (solving mathematical tasks) [Krämer et al. 2016] | **positive**; learners interacting with rapport-oriented PA have higher learning gains in a pre-/post-test |
| | | **Style of instruction**: PA using polite vs. direct language when formulating feedback and hints (mastering chemical stoichiometry) [McLaren et al. 2011] | **neutral**; no significant impact of PA's politeness on learning gains as measured by a pre-/post-test |
| | | **Style of instruction**: social robot – using the name of the learner, iconic gesture, polite language vs. asocial robot vs. no robot (learning about prime numbers) [Kennedy et al. 2015] | **mixed**; only the learning gains, i.e., difference between pre- and post-test, in the asocial condition were significantly high, in the other conditions no significant learning gains were reported |
| | | **Style of instruction**: virtual human featuring positive vs. negative emotional feedback in form of gesture (learning tai chi martial arts) [Bian et al. 2016] | **positive**; positive emotional feedback leads to higher learning outcomes as measured by expert assessment |
| | | **Social supportiveness**: responsive PA reacting verbally to learner's affect vs. non-responsive PA (designing an e-learning class for children) [Kim et al. 2007] | **neutral**; no significant impact of the degree of empathetic response on learners' learning as measured by a recall post-test |
| | | **Social supportiveness**: responsive PA – clapping hands, cheering, as reaction to learner' performance, vs. neutral style of interaction (memorizing multiplication table) [Konijn and Hoorn 2020] | **mixed**; socially supportive PA better enhanced the performance of above-average learners, while the neutral PA supported the enhanced performance of below-average learners |
| | | **Social supportiveness**: responsive PA – nods, shakes head, guides attention, etc. vs. neutral PA | **positive**; learners in the socially supportive condition outperform learners in the neutral condition in terms of language post-test scores |

| | | |
|---|---|---|
| | (learning foreign language) [Saerbeck et al. 2010] | |
| PAAS: Division of Labor | **Source of feedback**: PA-generated feedback using a template vs. manually inserted by a human tutor (translation to first order logic) [Perikos et al. 2017] | **neutral**; template-based automatically generated feedback and feedback manually inserted by the human tutor have comparable impact on the learning outcome |
| | **Pedagogical role of the PA**: peer – adaptively switching between tutor and tutee role, vs. fixed tutee – receiving instructions from the learner, vs. fixed tutor – providing instructions to the learner (vocabulary learning) [Chen et al. 2020] | **peer > tutor = tutee**; learners learned more words when interacting with PA acting as a peer, which switches between receiving and providing instructions based on learner's engagement, than learners who interacted with a fixed tutor or tutee PA |
| | **Pedagogical role of the PA**: cognitive – focused on facts and actions, vs. motivational – focused on emotions and feelings, vs. mixed (mastering formatting in MS Word) [van der Meij 2013] | **motivational > cognitive = mixed**; learners who interacted with the motivational PA outperformed learners in other conditions in terms of learning gain according to a pre-/post-test |
| | **Pedagogical role of the PA**: mentor – combining expert and motivator, vs. expert – focused on information provision, vs. motivator – focused on encouragement (creating instructional plan as a teacher) [Baylor and Kim 2005] | **mentor > expert > motivator**; learners in the mentor condition outperform others in an ability transfer post-test, learners in mentor and expert conditions outperform the motivator condition |

# References (Appendix only, see p. 42 for other references)